\documentclass[a4paper,fleqn,usenatbib]{mnras}

\usepackage{newtxtext,newtxmath}
\usepackage[T1]{fontenc}
\usepackage{ae,aecompl}
\usepackage{bm}

\usepackage{float}
\usepackage[export]{adjustbox}
\usepackage{graphicx}	% Including figure files
\usepackage{amsmath}	% Advanced maths commands
\usepackage{amssymb}	% Extra maths symbols

\usepackage{multicol}
\usepackage{verbatim}   % For commenting out multiple lines. --JR
\usepackage{threeparttable}
\usepackage{upgreek}

\newcommand{\lcdm}{$\Lambda$CDM}
\newcommand{\pcdm}{$\phi$CDM}
\newcommand{\om}{\Omega_{m0}}
\newcommand{\ol}{\Omega_{\Lambda}}
\newcommand{\ok}{\Omega_{k0}}
\newcommand{\FT}[1]{}

\title[Constraints from GRB data]{Constraints on cosmological parameters from gamma-ray burst peak photon energy and bolometric fluence measurements and other data}

\author[N. Khadka, B. Ratra]{
 Narayan Khadka,$^{1}$\thanks{E-mail: nkhadka@phys.ksu.edu}
and Bharat Ratra,$^{1}$\thanks{E-mail: ratra@phys.ksu.edu}
\\
$^{1}$Department of Physics, Kansas State University, 116 Cardwell Hall, Manhattan, KS 66502, USA\\
}

\date{Accepted XXX. Received YYY; in original form ZZZ}

\pubyear{2019}

\begin{document}
\label{firstpage}
\pagerange{\pageref{firstpage}--\pageref{lastpage}}
\maketitle

\begin{abstract}
We use measurements of the peak photon energy and bolometric fluence of 119 gamma-ray bursts (GRBs) extending over the redshift range of $0.3399 \leq z \leq 8.2$ to simultaneously determine cosmological and Amati relation parameters in six different cosmological models. The resulting Amati relation parameters are almost identical in all six cosmological models, thus validating the use of the Amati relation in standardizing these GRBs. The GRB data cosmological parameter constraints are consistent with, but significantly less restrictive than, those obtained from a joint analysis of baryon acoustic oscillation and Hubble parameter measurements.
%%%
\end{abstract}


%%%%%%%%%%%%%%%%%%%%%%%%%%%%%%%%%%%%%%%%%%%%%%%%%%

%%%%%%%%%%%%%%%%% BODY OF PAPER %%%%%%%%%%%%%%%%%%

\section{Introduction}
\label{sec:Introduction}
If general relativity provides an accurate description of cosmological gravitation, dark energy is needed to explain the observed accelerated expansion of the current universe. At the current epoch, dark energy is the major contributor to the energy budget of the universe. Most cosmological models are based on the cold dark matter (CDM) scenario named after the second largest contribution to the current cosmological energy budget. There are a variety of CDM models under discussion now, based on different dark energy models. \cite{peeble1984} proposed that Einstein's cosmological constant $\Lambda$ contributes a large part of the current energy budget of the universe. In this spatially flat model---which is consistent with many cosmological measurements \citep{alam, Farooq2017, Scolnic2018, Plank2018}---$\Lambda$ is responsible for the accelerated expansion of the universe. This is the simplest CDM model which is observationally consistent with the accelerated expansion of the universe. In this $\Lambda$CDM standard model, the spatially homogenous cosmological constant $\Lambda$ contributes $\sim 70\%$ of today's cosmological energy budget, the second most significant contributor being the cold dark matter which contributes $\sim 25\%$, and third place is occupied by the ordinary baryonic matter which contributes $\sim 5\%$.

Observational data however do not yet have sufficient precision to rule out extensions of the standard spatially-flat $\Lambda$CDM model. For example, dynamical dark energy \citep{peebles1988} that slowly varies in time and space remains observationally viable. Slightly non-flat spatial geometries are also not inconsistent with current observational constraints.\footnote{For observational constraints on spatial curvature see \cite{Farooq2015}, \cite{chen6}, \cite{Yu2016}, \cite{Rana2017}, \cite{Ooba2018a, Ooba2018b, Ooba2018c}, \cite{DESa}, \cite{Yu2018}, \cite{Park2018a, Park2018b, Park2018d, Park2018c, Park2019}, \cite{wei2018}, \cite{Xu2019}, \cite{Li2019}, \cite{Giambo2019}, \cite{Cole2019}, \cite{Eingorn2019}, \cite{Jesus2019}, \cite{Handley2019}, \cite{Wang2019}, \cite{Zhai2019}, \cite{Geng2020}, \cite{Kumar2020}, \cite{Efsta2020}, \cite{Di2020}, \cite{Gao2020} and references therein.} In this paper\textbf{,} we study the $\Lambda$CDM model as well as dynamical dark energy models, both spatially-flat and non-flat.

One of the main goals in cosmology now is to find the cosmological model that most accurately approximates the universe. A related important goal is to measure cosmological parameters precisely. To accomplish these goals require more and better data. Cosmological models are now largely tested in the redshift range $0 < z < 2.3$, with baryon acoustic oscillation (BAO) measurements providing the $z \sim 2.3$ constraints, and with cosmic microwave background (CMB) anisotropy data at $z \sim 1100$. There are only a few cosmological probes that access the $z \sim 2$ to $z \sim 1100$ part of the universe. These include HII starburst galaxies which reach to $z \sim 2.4$ \citep[and references therein]{Siegel2005, Mania2012, Gonzalez2019, Caok}, quasar angular size measurements which reach to $z \sim 2.7$ \citep[and references therein]{Gurvits1999, chen03, Cao2017, Ryan2019, Caok}, and quasar flux measurements that reach to $z \sim 5$ \citep[and references therein]{Risaliti2015, Risaliti2019, yang2019, Khadka2019, Khadka2020}.

Gamma-ray burst (GRBs) are another higher redshift probe of cosmology \citep[and references therein]{Lamb2000, Amati2002, Amati2008, SamushiaR2010, Demianski2011, Liu2015, Lin2016, Wang2016, Demianski2017, Demianski2019, Amati2019, Dirisa2019, Kumar2020, Montiel2020}. As a consequence of the enormous energy released during the burst, GRBs have been observed at least up to $z \sim 8.2$ \citep{Wang2016, Demianski2017, Demianski2019, Amati2019}. The cosmology of the $z \sim 5-8$ part of the universe is to date primarily accessed by GRBs. So if we can standardize GRBs this could help us study a very large part of the universe that has not yet been much explored.

There have been many attempts to standardize GRBs using phenomenological relations \citep[and references therein]{Amati2002, Ghirlanda2004, Liang2005}. One such relation is the non-linear Amati relation \citep{Amati2002} between the peak photon energy $E_p$ and the isotropic-equivalent radiated energy $E_{\rm iso}$ of a GRB. Some of the analyses assume a given current value of the non-relativistic matter density parameter $(\Omega_{m0})$ in an assumed cosmological model when calibrating the Amati relation \citep{Amati2008, Demianski2011, Dirisa2019}. These analyses result in GRB cosmological constraints that tend to more favor the assumed cosmological model. Some analyses use supernovae to calibrate the GRB data \citep{Kodama2008, Liang2008, Wang2016, Demianski2017}. Although this method is model-independent, supernovae systematics can affect the calibration process. Another model-independent calibration of the Amati relation has been done using the Hubble parameter $[H(z)]$ data \citep{Amati2019, Montiel2020, Marco2020}. These attempts to standardize GRBs through the Amati relation use some external factors. In this sense, the resulting constraints from the GRB data are not pure GRB constraints. We test this relation and use it to constrain cosmological parameters in six different cosmological models simultaneously. From our study of the Amati relation in six different cosmological models, we find, for the GRB data we study, that the parameters of the Amati relation are independent of the cosmological model we consider. This means that the Amati relation can standardize the GRBs we consider and so makes it possible to use them as a cosmological probe. Our demonstration of the cosmological-model-independence of the Amati relation is the most comprehensive to date. 

The GRB data we use have large error bars and so do not provide restrictive constraints on cosmological parameters. However, the GRB constraints are consistent with those we derive from BAO and $[H(z)]$-data and so we also perform joint analyses of the GRB + BAO + $H(z)$ data. Future improvements in GRB data should provide more restrictive constraints and help fill part of the observational data gap between the highest $z \sim 2.3$ BAO data and the $z \sim 1100$ CMB anisotropy data.

This paper is organized as follows. In Sec. 2 we describe the cosmological models that we study. In Sec. 3 we discuss the data that we use to constrain cosmological parameters in these models. In Sec. 4 we describe the methodology adopted for these analyses. In Sec. 5 we present our results, and we conclude in Sec. 6.

%%%
\section{Models}
\label{sec:models}
We work with six different general relativity dark energy cosmological models. Three are spatially-flat, the other three allow for non-zero spatial curvature. Each model is used to compute the luminosity distances of cosmological events at known redshifts, which can be used to predict observed quantities in terms of the parameters of the cosmological model. The luminosity distance depends on the cosmological expansion rate---the Hubble parameter---which is a function of redshift $z$ and the cosmological parameters of the model.

In the $\Lambda$CDM model the Hubble parameter is
\begin{equation}
\label{eq:friedLCDM}
    H(z) = H_0\sqrt{\Omega_{m0}(1 + z)^3 + \Omega_{k0}(1 + z)^2 + \Omega_{\Lambda}},
\end{equation}
where $\Omega_{k0}$ and $\Omega_{\Lambda}$ are the current values of the spatial curvature energy density and cosmological constant dark energy density parameters, and $H_0$ is the Hubble constant. The three energy density parameters are related as $\Omega_{m0} + \Omega_{k0} + \Omega_{\Lambda} =1$. In the $\Lambda$CDM model\textbf{,} $\Omega_{\Lambda}$ is a constant. In the spatially-flat cosmological models $\Omega_{k0}=0$. In the spatially-flat $\Lambda$CDM model, $\Omega_{m0}$ and $H_0$ are chosen to be the free parameters. In the spatially non-flat $\Lambda$CDM model, $\Omega_{m0}$, $\Omega_{\Lambda}$, and $H_0$ are taken to be the free parameters.

In the XCDM parametrization the Hubble parameter is
\begin{equation}
\label{eq:XCDM}
    H(z) = H_0\sqrt{\Omega_{m0}(1 + z)^3 + \Omega_{k0}(1 + z)^2 + \Omega_{X0}(1+z)^{3(1+\omega_X)}},
\end{equation}
where $\Omega_{X0}$ is the current value of the $X$-fluid dynamical dark energy density parameter and $\omega_X$ is the equation of state parameter for the $X$-fluid. The $X$-fluid pressure $P_X$ and energy density \textbf{$\rho_X$} are related through the equation of state $P_X = \omega_X \rho_X$. The three energy density parameters are related as $\Omega_{m0} + \Omega_{k0} + \Omega_{X0} = 1$. In the $\omega_X = -1$ limit the XCDM parametrization reduces to the $\Lambda$CDM model. In this parametrization dark energy is dynamical when $\omega_X \neq -1$, and when $0 > \omega_X > -1$ it's energy density decreases with time. In the spatially-flat XCDM parametrization, $\Omega_{m0}$, $\omega_X$, and $H_0$ are taken to be the free parameters while in the non-flat XCDM parametrization, $\Omega_{m0}$, $\Omega_{k0}$, $\omega_X$, and $H_0$ are chosen to be the free parameters.

In the $\phi$CDM model\textbf{,} dynamical dark energy is modeled as a scalar field $\phi$ \citep{peebles1988, Ratra1988, Pavlov2013}.\footnote{For observational constraints on the $\phi$CDM model see \cite{yashar2009}, \cite{Samushia2010}, \cite{camp}, \cite{Farooq2013b}, \cite{Farooq2013a}, \cite{Avsa}, \cite{Sola2017}, \cite{Sola2018, Sola2019}, \cite{Zhai2017}, \cite{Ooba2018b, Ooba2018d}, \cite{Sangwan2018}, \cite{Park2018a}, \cite{Singh2019}, \cite{Mitra2019b}, \cite{Caok}, and references therein.} Here we consider a scalar field potential energy density $V(\phi)$ of the inverse power law form
\begin{equation}
\label{eq:phiCDMV}
    V(\phi) = \frac{1}{2}\kappa m_{p}^2 \phi^{-\alpha},
\end{equation}
where $m_{p}$ is the Planck mass, $\alpha$ is a positive parameter, and $\kappa$ is a function of $\alpha$
\begin{equation}
\label{eq:kappa}
  \kappa = \frac{8}{3}\left(\frac{\alpha + 4}{\alpha + 2}\right)\left[\frac{2}{3}\alpha(\alpha + 2)\right]^{\alpha/2} .
\end{equation}
With this potential energy density, the equations of motion of a homogenous cosmological model are
\begin{equation}
\label{field}
    \ddot{\phi} + \frac{3\dot{a}}{a}\dot\phi - \frac{1}{2}\alpha \kappa m_{p}^2 \phi^{-\alpha - 1} = 0,
\end{equation}
and,
\begin{equation}
\label{friedpCDM}
    \left(\frac{\dot{a}}{a}\right)^2 = \frac{8 \uppi}{3 m_{p}^2}\left(\rho_m + \rho_{\phi}\right) - \frac{k}{a^2}.
\end{equation}
Here overdots denote derivatives with respect to time, $k$ is positive, zero, and negative for closed, flat, and open spatial hypersurfaces, $\rho_m$ is the non-relativistic matter density, and $\rho_{\phi}$ is the contribution to the energy density from the scalar field
\begin{equation}
    \rho_{\phi} = \frac{m^2_p}{32\pi}[\dot{\phi}^2 + \kappa m^2_p \phi^{-\alpha}].
\end{equation}
By solving eqs. (5) and (6) numerically we can compute the scalar field energy density parameter
\begin{equation}
    \Omega_{\phi}(z, \alpha) = \frac{8 \uppi \rho_{\phi}}{3 m^2_p H^2_0}.
\end{equation}
The Hubble parameter in the $\phi$CDM model is
\begin{equation}
    H(z) = H_0\sqrt{\Omega_{m0}\left(1 + z\right)^3 + \Omega_{k0}\left(1 + z\right)^2 + \Omega_{\phi}\left(z, \alpha\right)},
\end{equation}
where $\Omega_{m0} + \Omega_{k0} + \Omega_{\phi}(z = 0, \alpha) = 1$ and in the limit $\alpha\rightarrow0$ the $\phi$CDM model reduces to the $\Lambda$CDM model. In the spatially-flat $\phi$CDM model, $\Omega_{m0}$, $\alpha$, and $H_0$ are chosen to be the free parameters and in the non-flat $\phi$CDM model, $\Omega_{m0}$, $\Omega_{k0}$, $\alpha$, and $H_0$ are taken to be the free parameters.

%%%
%Old section
%%%
%%%
%end old section
%%%

\section{Data}
\label{sec:data}
%%%
In this paper, we use GRB, BAO, and $H(z)$ data to constrain cosmological model parameters. 

We use 25 GRB measurements (hereafter D19) from \cite{Dirisa2019} over the redshift range of $0.3399 \leq z \leq 4.35$, given in Table 2 of \cite{Dirisa2019}. We also use 94 GRB measurements (hereafter W16) from \cite{Wang2016} over the redshift range $0.48 \leq z \leq 8.2$, given in Table 5 of \cite{Dirisa2019}. The GRB measurements used in our analyses are $z$, peak photon energy $(E_p)$, and bolometric fluence $(S_{\rm bolo})$ with their corresponding 1$\sigma$ uncertainties.\footnote{The only non-zero $z$ error is that for GRB 080916C of D19. In the flat and non-flat $\Lambda$CDM models including or excluding this $z$ error in the analysis results in no noticeable difference, and so we ignore it in our analyses.} The D19 $S_{\rm bolo}$ data we use are those computed for the $1-10^4$ keV energy band (the F10 values). As discussed in the next section, the value of $S_{\rm bolo}$ and it's uncertainty can be used to obtain the isotropic radiated energy $(E_{\rm iso})$ and the uncertainty on $E_{\rm iso}$. 

For some GRBs,  $E_{\rm iso}$ and $E_p$ are empirically found to be related through the Amati relation \citep{Amati2002}, a non-linear relation between these observed quantities that is discussed in the next section. The use of GRB data for cosmological purposes are based on the validity of this relation. This relation has two free parameters and an intrinsic dispersion ($\sigma_{\rm ext}$). By simultaneously fitting to the Amati relation and cosmological parameters in six different cosmological models, we find that these Amati relation parameter values are almost independent of cosmological model. Values of $\sigma_{\rm ext}$ determined by D19 and W16 in the spatially-flat $\Lambda$CDM model are around 0.48 and 0.38 respectively. The value of $\sigma_{\rm ext}$ for D19 is higher than that obtained from W16. This is expected because D19 has only about a quarter the number of GRBs as does W16.

The BAO data we use are listed in Table 1 of \cite{Caok}. It includes 11 measurements extending over the redshift range $0.122 \leq z \leq 2.34$. The $H(z)$ data we use are listed in Table 2 of \cite{Ryan2018}. It includes 31 measurements extending over the redshift range $0.07 \leq z \leq 1.965$.

In this paper, we determine cosmological and Amati relation parameter constraints from the D19 and W16 GRB data. The Amati relation parameters obtained from these two GRB data sets are consistent with each other. So we also determine the constraints on the cosmological and Amati relation parameters using the combined D19 + W16 GRB data. The D19 + W16 data constraints are consistent with the BAO + $H(z)$ ones, so we jointly analyze these GRB data and the BAO + $H(z)$ data.

\begin{figure}
    \includegraphics[width=\linewidth,right]{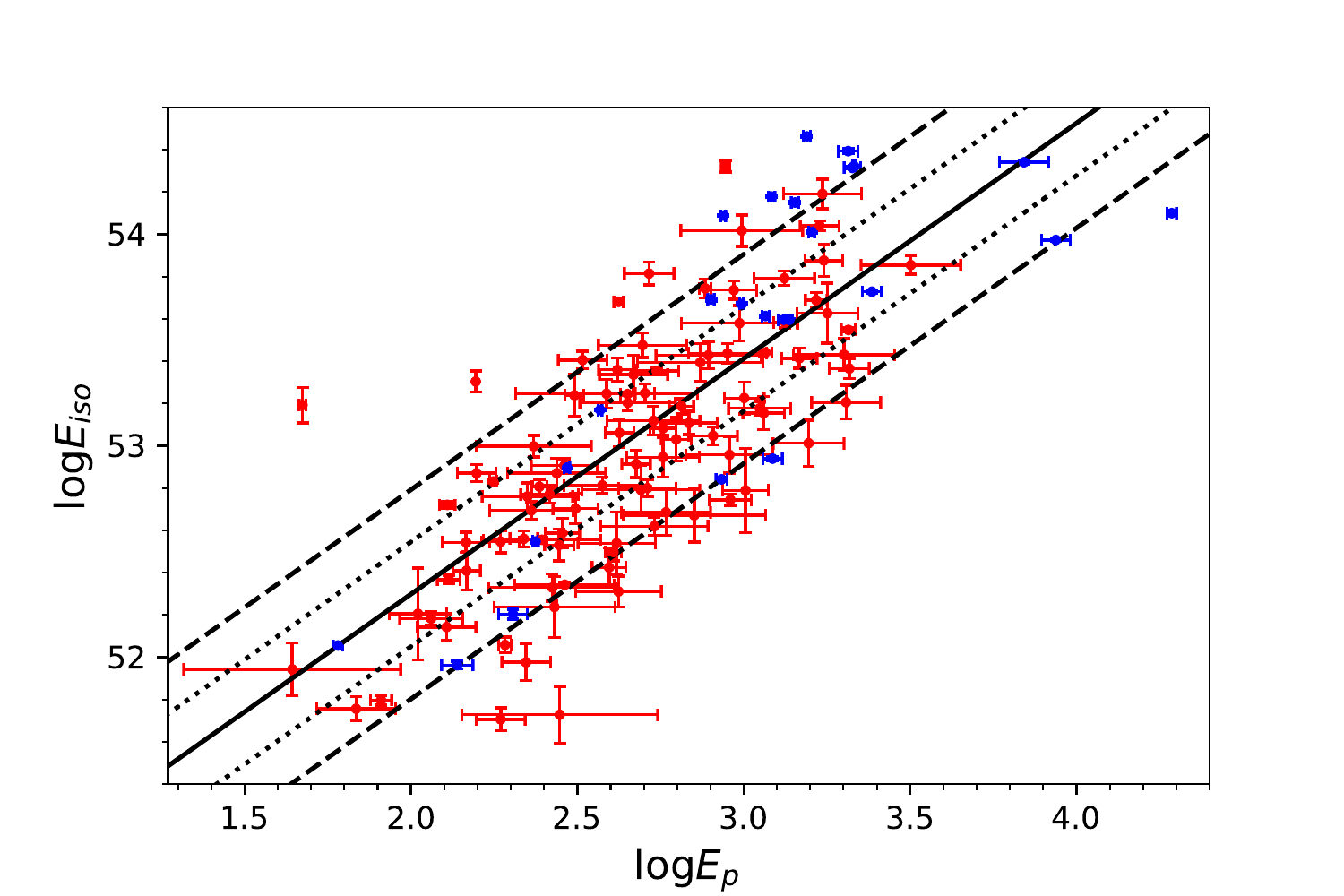}\par
\caption{$E_{\rm iso}-E_p$ correlation of the 119 GRBs for the spatially-flat $\Lambda$CDM model. Blue crosses show the 25 D19 GRB data and red crosses show the 94 W16 GRB data with 1$\sigma$ error bars. Black solid line is the Amati relation with best-fit parameter values, dotted and dashed lines are the Amati relations for the $\pm1\sigma$ and $\pm2\sigma$ values of the intercept $(a)$.}
\label{fig:Eiso-Ep}
\end{figure}

\section{Methods}
\label{sec:methods}
GRBs have been observed to high redshift, at least to $z = 8.2$. If it is possible to standardize GRBs, they can then be used as a cosmological probe to study a part of the universe which is not presently accessible to any other cosmological probe. For some GRBs the observed peak photon energy and isotropic energy are related through the Amati relation \citep{Amati2002}. Figure 1 shows that the GRB data we use here are related non-linearly through the Amati relation. The Pearson correlation coefficient between $\log(E_p)$ and $\log(E_{\rm iso})$ is 0.774 in the flat $\Lambda$CDM model. Given the quality of the data, this is a reasonably high correlation. This allows us to use these GRB data to constrain cosmological model parameters.

The Amati relation between the GRB's peak photon energy in the cosmological rest frame, $E_{p}$ and $E_{\rm iso}$ is given by
\begin{equation}
\label{eq:XCDM}
    \log(E_{\rm iso}) = a + b \log(E_{p}) ,
\end{equation}
where $\log = \log_{10}$, and $a$ and $b$ are the intercept and slope of the Amati relation and are free parameters to be determined from the data. Here $E_{p}$ and $E_{\rm iso}$ are defined through
\begin{equation}
\label{eq:XCDM}
    E_{\rm iso} = \frac{4\pi D^2_L(z, p) S_{\rm bolo}}{(1+z)} ,
\end{equation}

\begin{equation}
\label{eq:XCDM}
    E_{p} = E_{p,\rm obs} (1+z) ,
\end{equation}
where $S_{\rm bolo}$ is the measured bolometric fluence and $E_{ p,\rm obs}$ is the measured peak energy of the gamma-ray burst. Here the luminosity distance $D_L(z,p)$ is a function of redshift $z$ and cosmological parameters $p$ and is given by \citep{Khadka2019}
\begin{equation}
\label{eq:DM}
  \frac{H_0\sqrt{\left|\Omega_{k0}\right|}D_L(z, p)}{(1+z)} = 
    \begin{cases}
    {\rm sinh}\left[g(z)\right] & \text{if}\ \Omega_{k0} > 0, \\
    \vspace{1mm}
    g(z) & \text{if}\ \Omega_{k0} = 0,\\
    \vspace{1mm}
    {\rm sin}\left[g(z)\right] & \text{if}\ \Omega_{k0} < 0,
    \end{cases}   
\end{equation}
where
\begin{equation}
\label{eq:XCDM}
   g(z) = H_0\sqrt{\left|\Omega_{k0}\right|}\int^z_0 \frac{dz'}{H(z')},
\end{equation}
and $H(z)$ is the Hubble parameter listed in Sec. 2 for the cosmological models we use. GRB data cannot constrain $H_0$ because there is a degeneracy between the intercept parameter $a$ and $H_0$. So for GRB-only data analyses we set $H_0 = 70$ ${\rm km}\hspace{1mm}{\rm s}^{-1}{\rm Mpc}^{-1}$ but when using GRB data in conjuction with other data we allow $H_0$ to be a free parameter.

Using eqs. (10)--(12) together we can predict the bolometric fluence of a gamma-ray burst from $E_p$ and $z$, as a function of $D_L(z,p)$. We then compare these predicted values of bolometric fluence with the corresponding measured values using a likelihood function $({\rm LF})$. To avoid a circularity problem we fit the cosmological and Amati relation parameters simultaneously. The likelihood function we use for the GRB data is \citep{DAgostini2005}
\begin{equation}
\label{eq:chi2}
    \ln({\rm LF}) = -\frac{1}{2}\sum^{N}_{i = 1} \left[\frac{[\log(S^{\rm obs}_{\rm bolo,i}) - \log(S^{\rm th}_{\rm bolo,i})]^2}{s^2_i} + \ln(s^2_i)\right],
\end{equation}
%\Sig^2_{\log(S_{\rm bolo}} + b^2 \Sig^{2}_{\log(E_{p})}
where $\ln = \log_e$ and $s^2_i = \sigma^2_{\log(S_{\rm bolo,i})} + b^2 \sigma^2_{\log(E_{p,i})} + \sigma^2_{\rm ext}$. Here, $\sigma_{\log(S_{\rm bolo,i})}$ is the error in the measured value of $\log(S_{\rm bolo,i})$, $\sigma_{\log(E_{p,i})}$ is the error in $\log(E_{p,i})$, and $\sigma_{\rm ext}$ is the intrinsic dispersion of the Amati relation. $\sigma_{\log(S_{\rm bolo,i})}$ and $\sigma_{\log(E_{p,i})}$ are computed using the method of error propagation. We maximize this likelihood function and find best-fit values and errors of all the free parameters.

For the uncorrelated BAO and $H(z)$ data \textbf{\citep{alam, Ryan2018, Ryan2019}}, the likelihood function is
\begin{equation}
\label{eq:chi2}
    \ln({\rm LF}) = -\frac{1}{2}\sum^{N}_{i = 1} \frac{[A_{\rm obs}(z_i) - A_{\rm th}(z_i, p)]^2}{\sigma^2_i},
\end{equation}
where $A_{\rm obs}(z_i)$ and $A_{\rm th}(z_i, p)$ are the observed and model-predicted quantities at redshift $z_i$ and $\sigma_i$ is the uncertainty in the observed quantity.
For the correlated BAO data, the likelihood function is
\begin{equation}
\label{eq:chi2}
    \ln({\rm LF}) = -\frac{1}{2} [A_{\rm obs}(z_i) - A_{\rm th}(z_i, p)]^T \textbf{C}^{-1} [A_{\rm obs}(z_i) - A_{\rm th}(z_i, p)],
\end{equation}
For the BAO data from \cite{alam} the covariance matrix \textbf{C} is given in eq. (19) of \cite{Khadka2019} and for the BAO data from \cite{de2019} the covariance matrix is given in eq. (27) of \cite{Caok}.

In the BAO data analysis, the sound horizon $(r_s)$ is computed using the approximate formula \citep{Aubourg2015}
\begin{equation}
\label{eq:XCDM}
    r_s = \frac{55.154 \exp[-72.3(\Omega_{\nu0}h^2 + 0.0006)^2]}{(\Omega_{b0}h^2)^{0.12807} + (\Omega_{cb0}h^2)^{0.25351}} ,
\end{equation}
where $\Omega_{cb0} = \Omega_{b0} + \Omega_{c0} = \Omega_{m0} - \Omega_{\nu0}$. Here $\Omega_{b0}$, $\Omega_{c0}$, and $\Omega_{\nu0} = 0.0014$ \citep{Caok} are the CDM, baryonic, and neutrino energy density parameters at the present time, respectively, and $h = H_0/(100$ ${\rm km}\hspace{1mm}{\rm s}^{-1}{\rm Mpc}^{-1})$.

The maximization of the likelihood function in our analysis is done using the Markov chain Monte Carlo (MCMC) method as implemented in the emcee package \citep{Foreman2013} in Python 3.7. The convergence of a chain is confirmed using the \cite{Goodman2010} auto-correlation time (the chain should satisfy $N/50 \geq \tau$, where $N$ is the iteration number (size of the chain) and $\tau$ is the mean auto-correlation time). In our analysis we use flat priors for all free parameters, except in the GRB-only analyses where we set $H_0 = 70$ ${\rm km}\hspace{1mm}{\rm s}^{-1}{\rm Mpc}^{-1}$. The range of parameters over which the prior is non-zero are $0 \leq \om \leq 1$, $0 \leq \ol \leq 1.3$, $-0.7 \leq \Omega_{k0} \leq 0.7$ (and $-0.6 \leq k \leq 0.5$), $-5 \leq \omega_X \leq 5$ ($-20 \leq \omega_X \leq 20$ for the GRB-only data sets), $0 \leq \alpha \leq 3$, $0.45 \leq h \leq 1.0$, $-20 \leq \ln{\sigma_{\rm ext}} \leq 10$, $0 \leq b \leq 5$, and $0 \leq a \leq 300$.

To quantify the goodness of fit we compute the Akaike Information Criterion $(AIC)$ and the Bayes Information Criterion $(BIC)$ values for each cosmological model using eqs. (20) and (21) of \cite{Khadka2019}. The degree of freedom for each model is $\rm dof$ $= n-d$, where $n$ is the number of data points in the data set and $d$ is the number of free parameters in the model.

\begin{table*}
	\centering
	\small\addtolength{\tabcolsep}{-1pt}
	\caption{Unmarginalized best-fit parameters for all data sets.}
	\label{tab:BFP}
	\begin{threeparttable}
	\begin{tabular}{lcccccccccccccc} % four columns, alignment for each
		\hline
		Model & Data set & $\om$ & $\ol$ & $\ok$ & $\omega_{X}$ & $\alpha$ & $H_0$\tnote{a} & $\sigma_{\rm ext}$ & $a$ & $b$ & $\chi^2_{\rm min}$ & dof & $AIC$ & $BIC$\\
		\hline
		Flat \lcdm\ & B\tnote{b} & 0.314 & 0.686 & - & - & - & 68.515 & - & - & - & 20.737 & 40 & 24.737 & 28.212\\
		& D19 & 0.997 & 0.003 & - & - & - & - & 0.440 & 50.148 & 1.086 & 24.809 & 21 & 32.809 & 37.685\\
		& W16 & 0.303 & 0.697 & - & - & - & - & 0.380 & 50.349 & 1.064 & 93.028 & 90 & 101.028 & 111.201\\
		& GRB\tnote{c} & 0.878 & 0.122 & - & - & - & - & 0.402 & 50.003 & 1.103 & 117.659 &  115 & 125.659 & 136.775\\
		& GRB\tnote{c} + B\tnote{b} & 0.314 & 0.686 & - & - & - & 68.450 & 0.404 & 50.192 & 1.137 & 138.247 &  156 & 148.247 & 163.654\\
		\hline
		Non-flat \lcdm\ & B\tnote{b} & 0.308 & 0.644 & 0.048 & - & - & 67.534 & - & - & - & 20.452 & 39 & 26.452 & 31.665\\
		& D19 & 0.968 & 1.299 & - & - & - & - & 0.398 & 50.180 & 0.983 & 26.049 & 20 & 36.049 & 42.143\\
		& W16 & 0.481 & 0.012 & - & - & - & - & 0.380 & 50.180 & 1.076 & 92.322 & 89 & 102.322 & 115.039\\
		& GRB\tnote{c} & 0.723 & 0.022 & - & - & - & - & 0.400 & 50.016 & 1.117 & 117.984 & 114 & 127.984 & 141.880\\
		& GRB\tnote{c} + B\tnote{b} & 0.308 & 0.635 & - & - & - & 67.235 & 0.402 & 50.204& 1.137 & 138.980 & 155 & 150.980 & 169.468\\
		\hline
		Flat XCDM & B\tnote{b} & 0.319 & 0.681 & - & $-0.867$ & - & 65.850 & - & - & - & 19.504 & 39 & 25.504 & 30.717\\
		& D19 & 0.976 & - & - & $4.097$ & - & - & 0.380 & 50.610 & 0.737 & 23.969 & 20 & 33.969 & 40.063\\
		& W16 & 0.077 & - & - & $-0.229$ & - & - & 0.374 & 50.236 & 1.049 & 95.599 & 89 & 105.599 & 118.315\\
		& GRB\tnote{c} & 0.292 & - & - & $-0.183$ & - & - & 0.404 & 50.042 & 1.106 & 116.443 & 114 & 126.443 & 140.339\\
		& GRB\tnote{c} + B\tnote{b} & 0.321 & 0.679 & - & $-0.853$ & - & 65.524 & 0.406 & 50.222 & 1.130 & 135.929 & 155 & 147.929 & 166.418\\
		\hline
		Non-flat XCDM & B\tnote{b} & 0.327 & 0.831 & $-0.158$ & $-0.732$ & - & 65.995 & - & - & - & 18.386 & 38 & 26.386 & 33.337\\
		& D19 & 0.980 & - & $0.002$ & $4.560$ & - & - & 0.386 & 50.628 & 0.726 & 23.208 & 19 & 35.208 & 42.541\\
		& W16 & 0.812 & - & $0.434$ & $0.094$ & - & - & 0.378 & 50.168 & 1.079 & 93.003 & 88 & 105.003 & 120.263\\
		& GRB\tnote{c} & 0.905 & - & $0.529$ & $-1.272$ & - & - & 0.397 & 49.946 & 1.112 & 119.562 & 113 & 131.562 & 148.237\\
		& GRB\tnote{c} + B\tnote{b} & 0.326 & 0.816 & $-0.142$ & $-0.745$ & - & 66.121 & 0.407 & 50.175 & 1.134 & 134.217 & 154 & 148.217 & 169.786\\
		\hline
		Flat \pcdm\ & B\tnote{b} & 0.318 & 0.682 & - & - & 0.361 & 66.103 & - & - & - & 19.581 & 39 & 25.581 & 30.794\\
		& D19 & 0.999 & - & - & - & 1.825 & - & 0.378 & 50.643 & 0.911 & 25.570 & 20 & 35.570 & 41.664\\
		& W16 & 0.999 & - & - & - & 1.782 & - & 0.379 & 49.958 & 1.097 & 91.167 & 89 & 101.167 & 113.883\\
		& GRB\tnote{c} & 0.997 & - & - & - & 2.436 & - & 0.398 & 49.939 & 1.122 & 117.360 & 114 & 127.360 & 141.256\\
		& GRB\tnote{c} + B\tnote{b} & 0.321 & 0.679 & - & - &  0.416 & 65.793 & 0.402 & 50.218 & 1.132 & 138.156 & 155 & 150.156 & 168.645\\
		\hline
		Non-flat $\phi$CDM & B\tnote{b} & 0.322 & 0.832 & $-0.154$ & - & 0.935 & 66.391 & - & - & - & 18.545 & 38 & 26.545 & 33.496\\
		& D19 & 0.997 & - & 0.003 & - & 1.755 & - & 0.389 & 50.895 & 0.837 & 24.200 & 19 & 36.200 & 43.514\\
		& W16 & 0.992 & - & 0.007 & - & 1.451 & - & 0.381 & 50.193 & 1.005 & 90.820 & 88 & 102.820 & 118.080\\
		& GRB\tnote{c} & 0.978 & - & 0.018 & - & 2.072 & - & 0.396 & 49.957 & 1.114 & 118.192 & 113 & 130.192 & 145.452\\
		& GRB\tnote{c} + B\tnote{b} & 0.323 & 0.792 & $-0.115$ & - & 0.808 & 66.343 & 0.399 & 50.202 & 1.126 & 138.419 & 154 & 152.419 & 173.989\\
		 \hline
	\end{tabular}
    \begin{tablenotes}
    \item[a]${\rm km}\hspace{1mm}{\rm s}^{-1}{\rm Mpc}^{-1}$.
    \item[b]BAO + $H(z)$.
    \item[c]D19 + W16.
    \end{tablenotes}
    \end{threeparttable}
\end{table*}

\begin{figure*}
\begin{multicols}{2}
    \includegraphics[width=\linewidth]{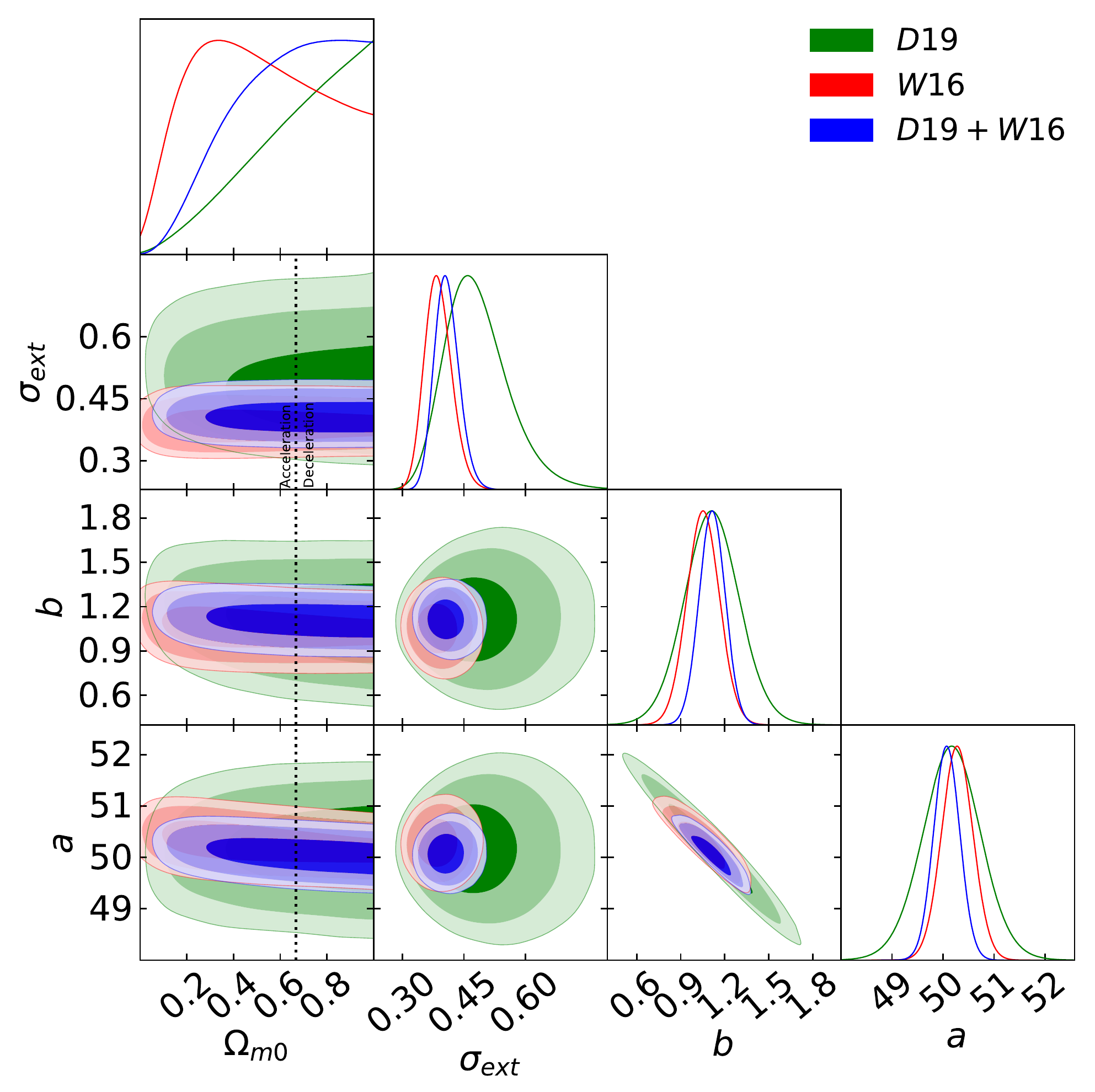}\par
    \includegraphics[width=\linewidth]{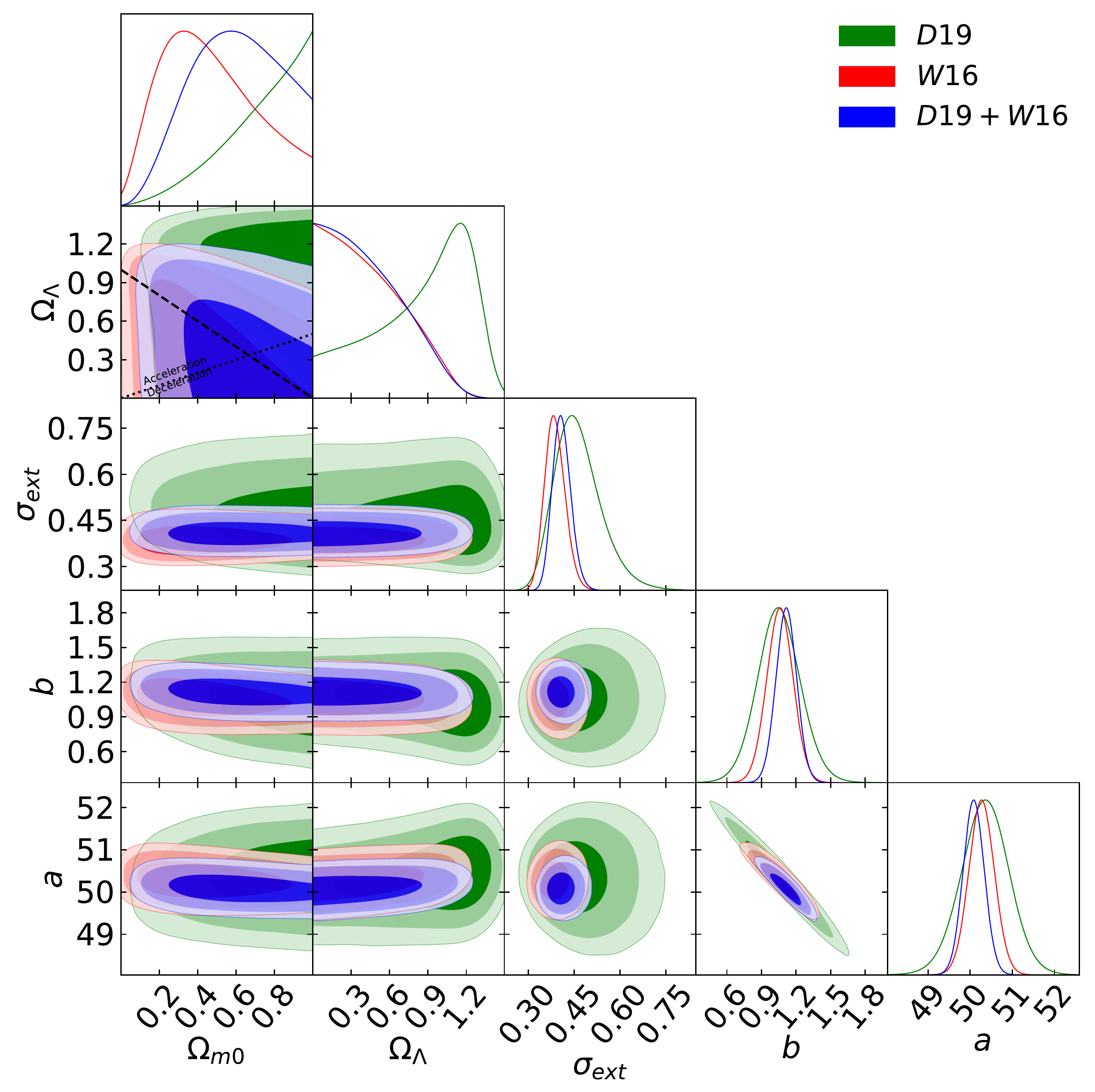}\par
\end{multicols}
\caption{One-dimensional likelihood distributions and two-dimensional contours at 1$\sigma$, 2$\sigma$, and 3$\sigma$ confidence levels using D19 (green), W16 (red), and D19 + W16 (blue) GRB data for all free parameters. Left panel shows the flat $\Lambda$CDM model. The black dotted lines are the zero acceleration line with currently accelerated cosmological expansion occurring to the left of the lines. Right panel shows the non-flat $\Lambda$CDM model. The black dotted lines in the $\Omega_{\Lambda}-\Omega_{m0}$ panel is the zero acceleration line with currently accelerated cosmological expansion occurring to the upper left of the line. The black dashed line in the $\Omega_{\Lambda}-\Omega_{m0}$ panel corresponds to the flat $\Lambda$CDM model, with closed hypersurface being to the upper right.}
\label{fig:flat LCDM68 model with BAO, H(z) and QSO data}
\end{figure*}

\begin{figure*}
\begin{multicols}{2}
    \includegraphics[width=\linewidth]{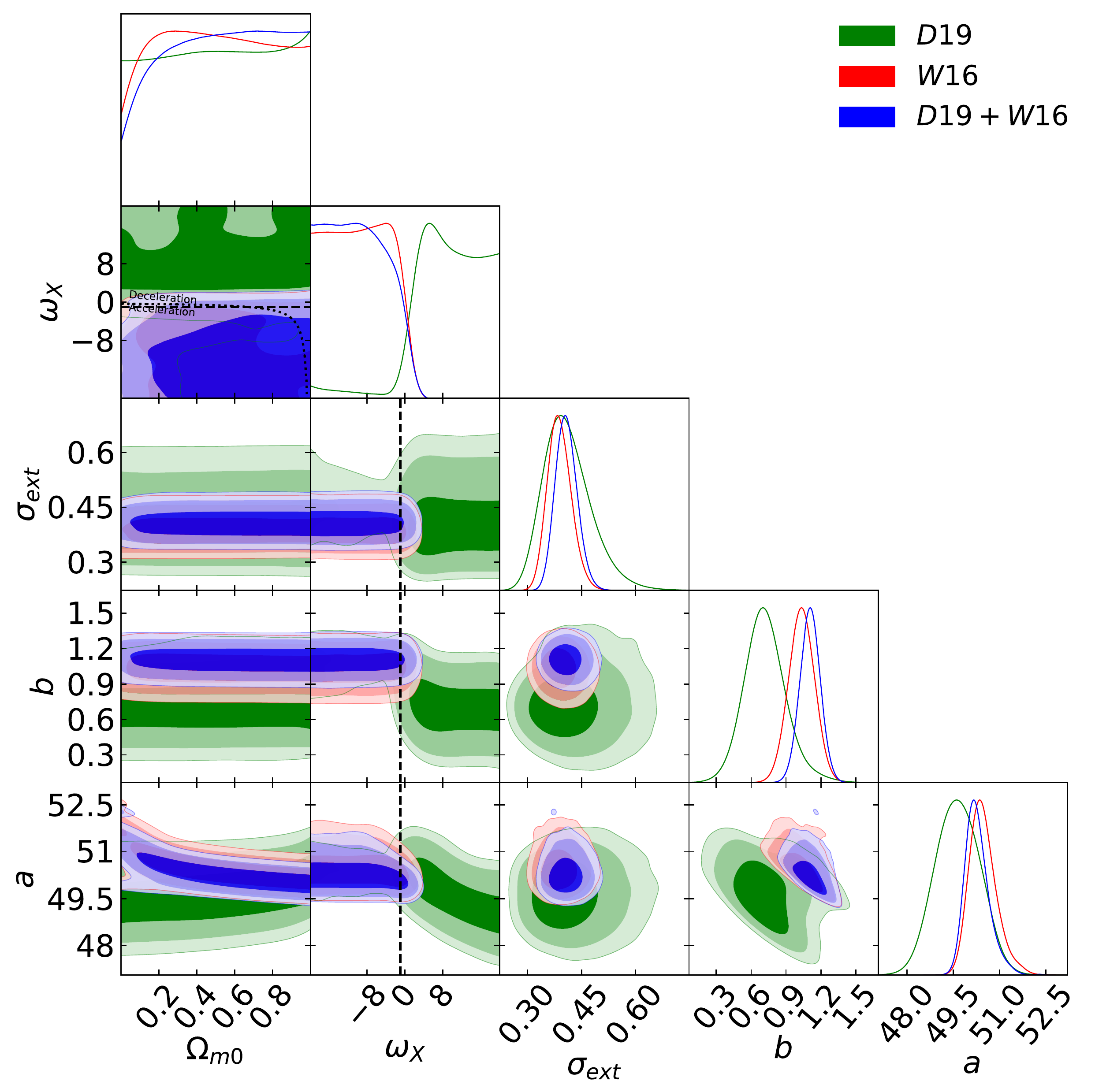}\par
    \includegraphics[width=\linewidth]{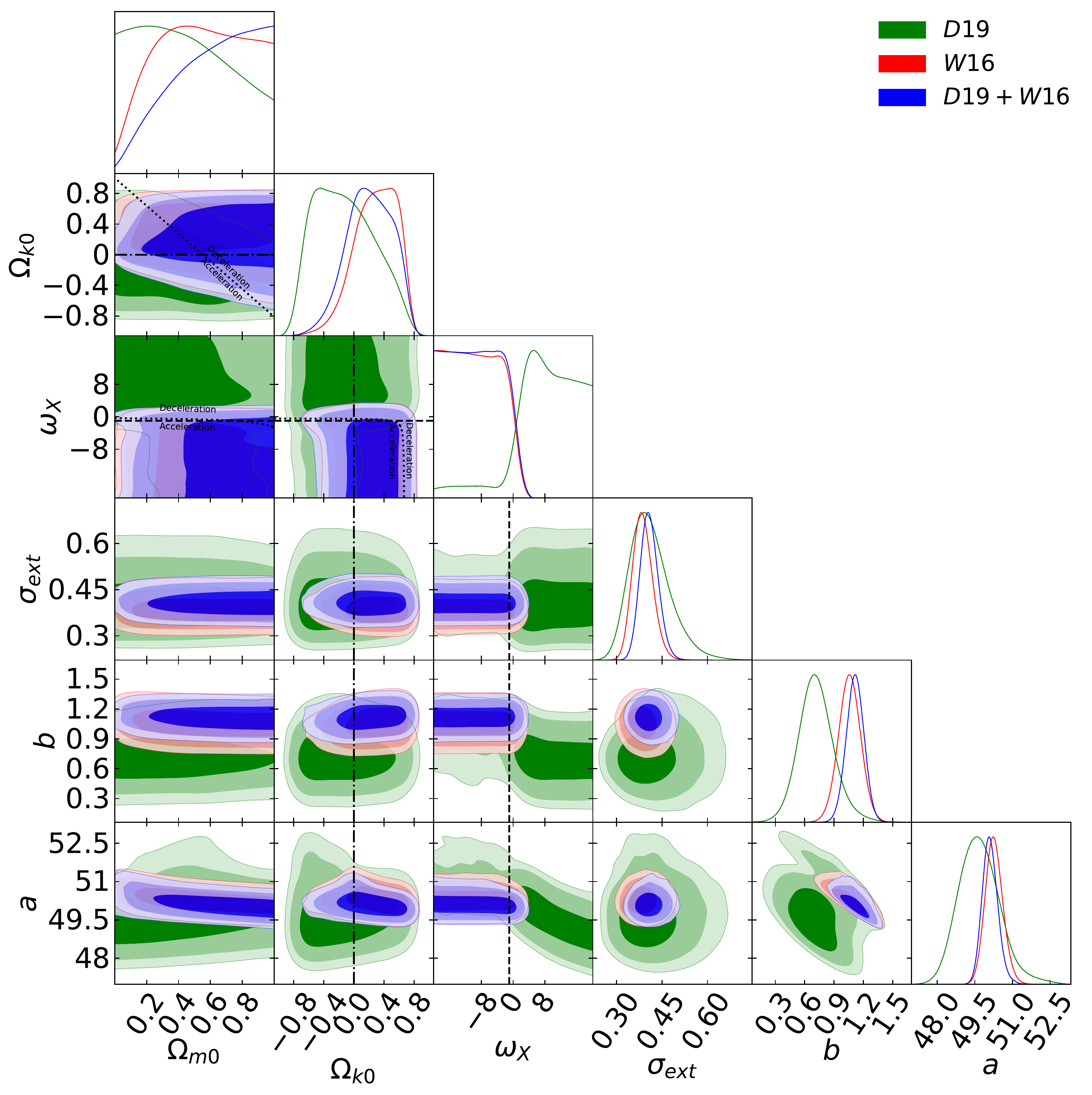}\par
\end{multicols}
\caption{One-dimensional likelihood distributions and two-dimensional contours at 1$\sigma$, 2$\sigma$, and 3$\sigma$ confidence levels using D19 (green), W16 (red), and D19 + W16 (blue) GRB data for all free parameters. Left panel shows the flat XCDM parametrization. The black dotted line in the $\omega_X-\Omega_{m0}$ panel is the zero acceleration line with currently accelerated cosmological expansion occurring below the line and the black dashed lines correspond to the $\omega_X = -1$ $\Lambda$CDM model. Right panel shows the non-flat XCDM parametrization. The black dotted lines in the $\Omega_{k0}-\Omega_{m0}$, $\omega_X-\Omega_{m0}$, and $\omega_X-\Omega_{k0}$ panels are the zero acceleration lines with currently accelerated cosmological expansion occurring below the lines. Each of the three lines \textbf{is} computed with the third parameter set to the GRB + BAO + $H(z)$ data best-fit value of Table 1. The black dashed lines correspond to the $\omega_x = -1$ $\Lambda$CDM model. The black dotted-dashed lines \textbf{correspond} to $\Omega_{k0} = 0$.}
\label{fig:flat LCDM68 model with BAO, H(z) and QSO data}
\end{figure*}

%\begin{figure*}
%   \includegraphics[width=\linewidth]{NLCDMgrb_f.pdf}\par
%\caption{NLCDM-GRBS}
%\label{fig:flat LCDM68 model with BAO, H(z) and QSO data}
%\end{figure*}

%\begin{figure*}
%    \includegraphics[width=\linewidth]{NXCDMgrbC1.pdf}\par
%\caption{NXCDM-GRBS}
%\label{fig:flat LCDM68 model with BAO, H(z) and QSO data}
%\end{figure*}

\begin{figure*}
\begin{multicols}{2}
    \includegraphics[width=\linewidth]{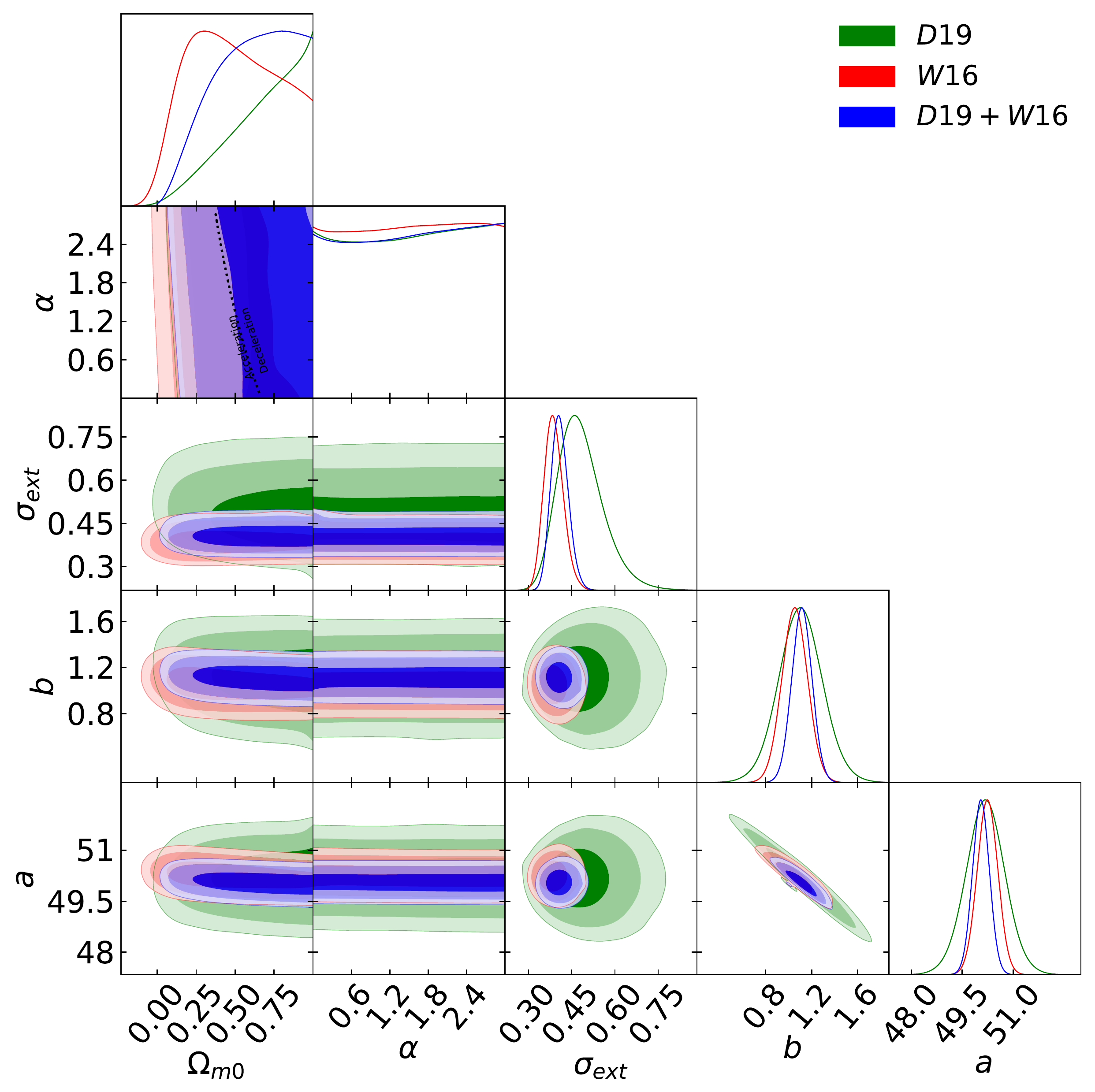}\par
    \includegraphics[width=\linewidth]{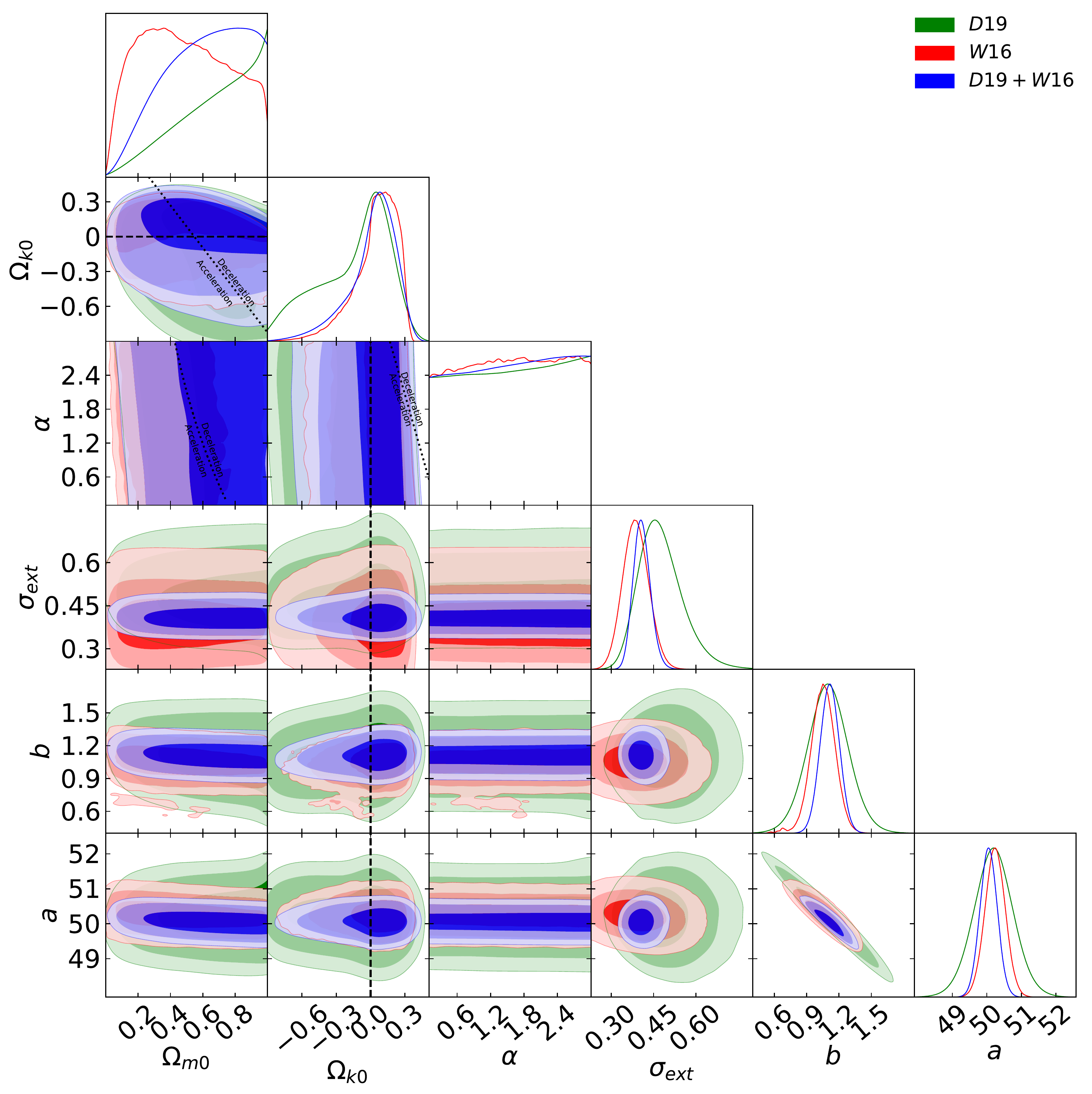}\par
\end{multicols}
\caption{One-dimensional likelihood distributions and two-dimensional contours at 1$\sigma$, 2$\sigma$, and 3$\sigma$ confidence levels using D19 (green), W16 (red), and D19 + W16 (blue) GRB data for all free parameters. Left panel shows the flat $\phi$CDM model. The black dotted curved line in the $\alpha - \Omega_{m0}$ panel is the zero acceleration line with currently accelerated cosmological expansion occurring to the left of the line. Right panel shows the non-flat $\phi$CDM model. The black dotted lines in the $\Omega_{k0}-\Omega_{m0}$, $\alpha-\Omega_{m0}$, and $\alpha-\Omega_{k0}$ panels are the zero acceleration lines with currently accelerated cosmological expansion occurring below the lines. Each of the three lines \textbf{is} computed with the third parameter set to the GRB + BAO + $H(z)$ data best-fit value of Table 1. The black dashed straight lines correspond to $\Omega_{k0} = 0$.}
\label{fig:flat LCDM68 model with BAO, H(z) and QSO data}
\end{figure*}

\begin{table*}
	\centering
	\small\addtolength{\tabcolsep}{-2.5pt}
	\small
	\caption{Marginalized one-dimensional best-fit parameters with 1$\sigma$ confidence intervals for all data sets. A 2$\sigma$ limit is given when only an upper or lower limit exists.}
	\label{tab:BFP}
	\begin{threeparttable}
	\begin{tabular}{lccccccccccc} % four columns, alignment for each
		\hline
		Model & Data set\hspace{5mm} & $\om$ & $\ol$ & $\ok$ & $\omega_{X}$ & $\alpha$ & $H_0$\tnote{a} & $\sigma_{ext}$ & $a$ & $b$ \\
		\hline
		Flat \lcdm\ & B\tnote{b} & $0.315^{+0.016}_{-0.016}$ & $0.685^{0.016}_{0.016}$ & - & - & - & $68.517^{+0.869}_{-0.869}$ & - & - & -\\
		& D19 & $> 0.269$ & $< 0.731$ & - & - & - & - & $0.475^{+0.085}_{-0.064}$ & $50.190^{+0.543}_{-0.560}$ & $1.109^{+0.181}_{-0.181}$\\
		& W19 & $> 0.125$ & $< 0.875$ & - & - & - & - & $0.386^{+0.034}_{-0.030}$ & $50.306^{+0.298}_{-0.303}$ & $1.052^{+0.109}_{-0.108}$\\
		& GRB\tnote{c} & $> 0.247$ & $< 0.753$ & - & - & - & - & $0.407^{+0.031}_{-0.027}$ & $50.070^{-0.247}_{-0.248}$ & $1.114^{+0.086}_{-0.087}$\\
		& GRB\tnote{c} + B\tnote{b} & $0.316^{+0.016}_{-0.016}$ & $0.684^{+0.016}_{-0.016}$ & - & - & - & $68.544^{+0.871}_{-0.862}$ & $0.409^{+0.029}_{-0.027}$ & $50.196^{+0.231}_{-0.230}$ & $1.134^{+0.083}_{-0.083}$\\
		\hline
		Non-flat \lcdm\ & B\tnote{b} & $0.309^{+0.016}_{-0.016}$ & $0.640^{+0.073}_{-0.077}$ & $0.051^{+0.095}_{-0.089}$ & - & - &$67.468^{+2.336}_{-2.311}$& - & - & -\\
		& D19 & $> 0.326$ & ---- & ---- & - & - & - & $0.457^{+0.084}_{-0.063}$ & $50.344^{+0.538}_{-0.563}$ & $1.051^{+0.182}_{-0.178}$\\
		& W16 & $0.432^{+0.299}_{-0.217}$ & $< 0.946$ & $> -0.976$ & - & - & - & $0.387^{+0.035}_{-0.030}$ & $50.269^{+0.297}_{-0.299}$ & $1.058^{+0.110}_{+0.113}$\\
		& GRB\tnote{c} & $0.596^{+0.249}_{-0.237}$ & $< 0.933$ & $> -1.027$ & - & - & - & $0.409^{+0.032}_{-0.028}$ & $50.081^{+0.245}_{-0.246}$ & $1.115^{+0.087}_{-0.088}$\\
		& GRB\tnote{c} + B\tnote{b} & $0.310^{+0.016}_{-0.016}$ & $0.639^{+0.072}_{-0.078}$ & $0.051^{+0.094}_{-0.088}$ & - & - &$67.499^{+2.281}_{-2.279}$ & $0.408^{+0.030}_{-0.027}$ & $50.203^{+0.234}_{-0.231}$ & $1.137^{+0.083}_{-0.084}$\\
		\hline
		Flat XCDM & B\tnote{b} & $0.319^{+0.017}_{-0.016}$ & $0.681^{+0.016}_{-0.017}$ & - & $-0.882^{+0.106}_{-0.121}$ & - &$66.185^{+2.575}_{-2.375}$&-&-&-\\
		& D19 & ---- & - & - & $> -2.793$ & - & - & $0.407^{+0.074}_{-0.056}$ & $49.648^{+0.724}_{-0.736}$ & $0.710^{+0.174}_{-0.159}$\\
		& W16 & ---- & - & - & $< 5.303$ & - & - & $0.388^{+0.036}_{-0.030}$ & $50.428^{+0.484}_{-0.383}$ & $1.033^{+0.106}_{-0.108}$\\
		& GRB\tnote{c} & ---- & - & - & $< 3.244$ & - & - & $0.407^{+0.032}_{-0.027}$ & $50.232^{+0.424}_{-0.323}$ & $1.103^{+0.085}_{-0.086}$\\
		& GRB\tnote{c} + B\tnote{b} & $0.321^{+0.017}_{-0.016}$ & $0.679^{+0.016}_{-0.017}$ & - & $-0.874^{+0.107}_{-0.121}$ & - &$66.058^{+2.557}_{-2.391}$& $0.409^{+0.031}_{-0.027}$ & $50.206^{+0.233}_{-0.237}$ & $1.132^{+0.085}_{-0.084}$\\
		\hline
		Non-flat XCDM & B\tnote{b} & $0.323^{+0.020}_{-0.021}$ & - & $-0.095^{+0.165}_{-0.177}$ & $-0.777^{+0.119}_{-0.202}$ &-& $66.171^{+2.477}_{-2.348}$ &-&-&-\\
		& D19 & ---- & - & $-0.145^{+0.457}_{-0.384}$ & $> -3.623$ & - & - & $0.406^{+0.075}_{-0.055}$ & $49.647^{+0.818}_{-0.758}$ & $0.714^{+0.183}_{-0.160}$\\
		& W16 & $> 0.125$ & - & $0.297^{+0.273}_{-0.300}$ & $< 5.104$ & - & - & $0.386^{+0.035}_{-0.030}$ & $50.249^{+0.353}_{-0.339}$ & $1.056^{+0.111}_{-0.111}$\\
		& GRB\tnote{c} & $> 0.202$ & - & $0.199^{+0.321}_{-0.306}$ & $< 5.442$ & - & - & $0.408^{+0.033}_{-0.028}$ & $50.075^{+0.306}_{-0.297}$ & $1.114^{+0.088}_{-0.090}$\\
		& GRB\tnote{c} + B\tnote{b} & $0.324^{+0.020}_{-0.020}$ & - & $-0.090^{+0.156}_{-0.161}$ & $-0.774^{+0.114}_{-0.193}$ & - &$66.002^{+2.491}_{-2.323}$ & $0.408^{+0.030}_{-0.027}$ & $50.198^{+0.234}_{-0.231}$ & $1.127^{+0.084}_{-0.084}$\\
		\hline
		Flat \pcdm\ & B\tnote{b} & $0.319^{+0.017}_{-0.016}$ & $0.681^{+0.016}_{-0.017}$ & - & - & $0.540^{+0.170}_{-0.490}$& $65.300^{+2.300}_{-1.800}$&-&-&-\\
		& D19 & $> 0.244$ & - & - & - & ---- & - & $0.474^{+0.082}_{-0.064}$ & $50.183^{+0.542}_{-0.542}$ & $1.102^{+0.181}_{-0.179}$\\
		& W16 & $> 0.101$ & - & - & - & ---- & - & $0.386^{+0.034}_{-0.030}$ & $50.245^{+0.290}_{-0.288}$ & $1.053^{+0.107}_{-0.106}$\\
		& GRB\tnote{c} & $> 0.210$ & - & - & - & ---- & - & $0.407^{+0.030}_{-0.027}$ & $50.052^{+0.238}_{-0.236}$ & $1.115^{+0.084}_{-0.084}$\\
		& GRB\tnote{c} + B\tnote{b} & $0.321^{+0.017}_{-0.017}$ & $0.679^{+0.017}_{-0.017}$ & - & - & $0.570^{+0.200}_{-0.500}$ & $65.200^{+2.300}_{-1.900}$ & $0.409^{+0.027}_{-0.030}$ & $50.215^{+0.232}_{-0.232}$ & $1.131^{+0.084}_{-0.084}$\\
		\hline
		Non-flat $\phi$CDM & B\tnote{b} & $0.321^{+0.017}_{-0.017}$ & - & $-0.130^{+0.160}_{-0.130}$ & - & $0.940^{+0.450}_{-0.650}$ & $65.900^{+2.300}_{-2.300}$ &-&-&-\\
		& D19 & > 0.236 & - & $-0.05^{+0.180}_{+0.471}$ & - & ---- & - & $0.470^{+0.083}_{-0.064}$ & $50.183^{+0.544}_{-0.562}$ & $1.100^{+0.182}_{-0.179}$\\
		& W16 & $0.473^{+0.326}_{-0.272}$ & - & $0.076^{+0.147}_{-0.209}$ & - & ---- & - & $0.387^{+0.036}_{-0.030}$ & $50.230^{+0.293}_{-0.306}$ & $1.054^{+0.111}_{-0.112}$ \\
		& GRB\tnote{c} & > 0.209 & - & $0.054^{+0.146}_{-0.235}$ & - & ---- & - & $0.408^{+0.031}_{-0.027}$ & $50.047^{+0.241}_{-0.244}$ & $1.116^{+0.087}_{-0.085}$\\
		& GRB\tnote{c} + B\tnote{b} & $0.321^{+0.017}_{-0.017}$ & - & $-0.120^{+0.150}_{-0.130}$ & - & $0.940^{+0.460}_{-0.630}$ & $65.800^{+2.300}_{-2.300}$ & $0.408^{+0.030}_{-0.027}$ & $50.202^{+0.232}_{-0.233}$ & $1.126^{+0.084}_{-0.084}$\\
		\hline
	\end{tabular}
    \begin{tablenotes}
    \item[a]${\rm km}\hspace{1mm}{\rm s}^{-1}{\rm Mpc}^{-1}$.
    \item[b]BAO + $H(z)$.
    \item[c]D19 + W16.
    \end{tablenotes}
    \end{threeparttable}
\end{table*}

%%Note A

\section{Results}
\label{sec:Results}
\subsection{D19, W16, and D19 + W16 GRB data constraints}
\label{sec:GRB}
The unmarginalized and marginalised best-fit values and $1\sigma$ uncertainties ($2\sigma$ limit when only an upper or lower limit exists) for all free parameters determined using GRB data sets are given in Tables 1 and 2 respectively.
One-dimensional likelihood distributions and two-dimensional constraint contours obtained using GRB data are shown in Figs.\ 2--4. In these figures the D19, W16, and D19 + W16 GRB data results are shown in green, red, and blue respectively. Use of GRB data to constrain cosmological model parameters is based on the assumption that the Amati relation is valid for the GRB data. Here we use these GRB data and simultaneously determine Amati relation parameters for six different cosmological models. This is the most comprehensive test of the Amati relation for a GRB data set done to date.

The Amati relation parameters for the three different GRB data sets are largely consistent with each other. In the flat $\Lambda$CDM model the slope parameter $(b)$ for the D19, W16, and D19 + W16 data sets is found to be $1.109^{+0.181}_{-0.181}$, $1.052^{+0.109}_{-0.108}$, and $1.114^{+0.086}_{-0.087}$, and the intercept parameter $(a)$ is found to be $50.190^{+0.543}_{-0.056}$, $50.306^{+0.298}_{-0.303}$, and $50.070^{+0.247}_{-0.248}$, respectively. In the non-flat $\Lambda$CDM model, for the D19, W16, and D19 + W16 data sets $b$ is found to be $1.051^{+0.182}_{-0.178}$, $1.058^{+0.110}_{-0.113}$, and $1.115^{+0.087}_{-0.088}$, and $a$ is found to be $50.344^{+0.538}_{-0.056}$, $50.269^{+0.297}_{-0.299}$, and $50.081^{+0.245}_{-0.246}$, respectively. Similar results hold for the flat and non-flat XCDM and $\phi$CDM cases. For the D19 data the measured values of $b$ are slightly lower in both the XCDM cases compared to the other models and GRB data sets. The Amati relation parameters for different data sets and cosmological models differ only slightly from each other. These differences between values for different GRB data sets are not unexpected because each data set has a different number of GRBs. For the combined D19 + W16 data these parameters are essentially independent of cosmological model.

Another free parameter which characterizes how well the GRB data can constrain cosmological model parameters is the intrinsic dispersion of the Amati relation $(\sigma_{\rm ext})$. In the flat $\Lambda$CDM model $\sigma_{\rm ext}$ for the D19, W16, and D19 + W16 data sets  is found to be $0.475^{+0.085}_{-0.064}$, $0.386^{+0.034}_{-0.030}$, and $0.407^{+0.031}_{-0.027}$, respectively. In the non-flat $\Lambda$CDM model for the D19, W16, and D19 + W16 data sets $\sigma_{\rm ext}$ is found to be $0.475^{+0.084}_{-0.063}$, $0.387^{+0.035}_{-0.030}$, and $0.409^{+0.032}_{-0.028}$ respectively. Similar results hold for the flat and non-flat XCDM and $\phi$CDM cases. The measured values of $\sigma_{\rm ext}$ are almost model-independent, especially for the combined D19 + W16 GRB data, which indicates that these data behave in the same way for all cosmological models considered here. The model-independent behavior of the Amati relation parameters and intrinsic dispersion signifies that these GRBs can be used as standard candles although, given the large error bars, they do not restrictively constrain cosmological parameters.

From Figs.\ 2--4 we see that for the combined D19 + W16 GRB data set, in most models currently accelerated cosmological expansion is more consistent with the observational constraints; a notable exception is the flat $\phi$CDM model, left panel of Fig.\ 4, where currently decelerated cosmological expansion is more favored.

Values of the non-relativistic matter density parameter, $\Omega_{m0}$, obtained using GRB data are given in Table 2. In the flat $\Lambda$CDM model, for the D19, W16, and D19 + W16 data $\Omega_{m0}$ is determined to be > 0.269, > 0.125, and > 0.247, respectively. In the non-flat $\Lambda$CDM model, for the D19, W16, and D19 + W16 data $\Omega_{m0}$ > 0.326, $= 0.432^{+0.299}_{-0.217}$, and $= 0.596^{+0.249}_{-0.237}$, respectively. In the flat XCDM parametrization, none of the three GRB data sets constrain $\Omega_{m0}$. In the non-flat XCDM case, the D19 data do not constrain $\Omega_{m0}$, and for the W16, and D19 + W16 data sets $\Omega_{m0}$  > 0.125, and > 0.202, respectively. In the flat $\phi$CDM model, for the D19, W16, and D19 + W16 data $\Omega_{m0}$  > 0.244, > 0.101, and > 0.210, respectively. In the non-flat $\phi$CDM model, for the D19, W16, and D19 + W16 data $\Omega_{m0}$ > 0.236, $= 0.473^{+0.326}_{-0.272}$, and > 0.209, respectively. These GRB data only weakly constrain and mostly provide a lower $2\sigma$ limit on $\Omega_{m0}$. These $\Omega_{m0}$ constraints are largely consistent with those determined using other data.

Values of the curvature energy density parameter $(\Omega_{k0})$ determined using GRB data sets are given in Table 2. In the non-flat $\Lambda$CDM model,\footnote{In the non-flat $\Lambda$CDM case, values of $\Omega_{k0}$ are computed (if possible) using the measured values of $\Omega_{m0}$ and $\Omega_{\Lambda}$ and the equation $\Omega_{m0} + \Omega_{k0} + \Omega_{\Lambda} = 1$.} the D19 data cannot constrain $\Omega_{k0}$ and for the W19 and D19 + W19 data $\Omega_{k0}$ is $> -0.976$, and $> -1.027$, respectively. In the non-flat XCDM parametrization, for the D19, W16, and D19 + W16 data $\Omega_{k0}$ is $= -0.145^{+0.457}_{-0.384}$, $= 0.297^{+0.273}_{-0.300}$, and $= 0.199^{+0.321}_{-0.306}$, respectively. In the non-flat $\phi$CDM model, for the D19, W16, and D19 + W16 data $\Omega_{k0}$ is $-0.050^{+0.180}_{-0.471}$, $0.076^{+0.147}_{-0.209}$, and $0.054^{+0.146}_{-0.235}$, respectively.

In the flat $\Lambda$CDM model, for the D19, W16, and D19 + W16 data sets the cosmological constant energy density parameter $(\Omega_{\Lambda})$ is measured to be < 0.731, < 0.875, and < 0.753, respectively. In the non-flat $\Lambda$CDM model the D19 data cannot constrain $\Omega_{\Lambda}$ at the 2$\sigma$ confidence level and values of $\Omega_{\Lambda}$ for the W16 and D19+W16 data sets are found to be < 0.946, and < 0.933, respectively.

In the flat (non-flat) XCDM parametrization, for the D19, W16, and D19 + W16 data the equation of state parameter $(\omega_X)$ is determined to be $> - 2.793 (> -3.623)$, $< 5.303 (< 5.104)$, and $< 3.244 (< 5.442)$ respectively. None of the GRB data sets are able to constrain the scalar field potential energy density parameter $\alpha$ of the $\phi$CDM model.

From the values of $AIC$, and $BIC$ listed in Table 1, The most favored model for all three GRB data sets is the flat $\Lambda$CDM model. The least favored case for the D19 + W16 data is the non-flat XCDM parametrization.

\subsection{Constraints from BAO + $H(z)$ data}
\label{sec:Constraints from the BAO + $H(z)$}
The BAO data that we use here \textbf{is} an updated compilation compared to what we used in \cite{Khadka2020}. These were first used in \cite{Caok} although the BAO + $H(z)$-only results were not shown or discussed there. Unmarginalized and marginalized best-fit values of all free parameters are given in Tables 1 and 2. One-dimensional likelihood distributions and two-dimensional constraint contours are shown in red in Figs.\ 5--10. 

From Table 2, for the BAO + $H(z)$ data the value of the non-relativistic matter density parameter $(\Omega_{m0})$ ranges from $0.309 \pm 0.016$ to $0.323^{+0.020}_{-0.021}$. The lowest value is obtained in the non-flat $\Lambda$CDM model and the highest value in the non-flat XCDM parametrization. 

We can also constrain the Hubble constant using BAO + $H(z)$ data. We find that the Hubble constant $(H_0)$ ranges from $65.300^{+2.300}_{-1.800}$ to $68.517 \pm 0.869$ ${\rm km}\hspace{1mm}{\rm s}^{-1}{\rm Mpc}^{-1}$. The lowest value is obtained for the spatially-flat $\phi$CDM model and the highest value for the spatially-flat $\Lambda$CDM model. These values are more consistent with the \cite{Plank2018} result than with the local expansion rate measurement of \cite{Riess2016}.\footnote{They are also consistent with median statistics estimates \citep{Gott2001, chen1, chen03} and a number of recent measurements \citep{chen5, Zhang2017,  Dhaw, Fernandez2018, DESb, Yu2018, Gomez2018, Haridasu2018, Zhang2018a, D, Martinelli2019, Cuceu2019, Freedman2019, Freedman2020, Zeng2019, schon2019,  LinI2019, Rameez2019, Zhang2019, phil2020}.}

Values of curvature energy density parameter $(\Omega_{k0})$ determined using BAO + $H(z)$ data sets are given in Table 2. In the non-flat $\Lambda$CDM model (see footnote 4) $\Omega_{k0}$ is $0.051^{+0.095}_{-0.089}$. In the non-flat XCDM parametrization and $\phi$CDM model $\Omega_{k0}$ is $-0.095^{+0.165}_{-0.177}$, and $-0.130^{+0.160}_{-0.130}$, respectively. 

The value of the cosmological constant energy density parameter $(\Omega_{\Lambda})$ for the flat (non-flat) $\Lambda$CDM model is determined to be $0.685 \pm 0.016 (0.640^{+0.073}_{-0.079})$.

The equation of state parameter $(\omega_X)$ of the flat (non-flat) XCDM parametrization is measured to be $\omega_X = -0.882^{+0.106}_{-0.121} (-0.777^{+0.119}_{-0.202})$. The value of the scalar field potential energy density parameter $(\alpha)$ of the flat (non-flat) $\phi$CDM model is measured to be $\alpha = 0.540^{+0.170}_{-0.490} (0.940^{+0.450}_{-0.650})$. Measurements of both parameters favor dynamical dark energy.

From the $AIC$ and $BIC$ values listed in Table 1, the most favored model for the BAO + $H(z)$ data is the flat $\Lambda$CDM model and the least favored is the non-flat $\phi$CDM model.

\subsection{Constraints from GRB + BAO + $H(z)$ data}
\label{sec:Constraints from the GRB + BAO + $H(z)$}
Constraints obtained from the GRB data, are not very restrictive but are consistent with those obtained from the BAO + $H(z)$ data so it is reasonable to do joint analyses of these data. The sum of eqs. \ (15), (16), and (17) gives the $\ln({\rm LF})$ for the joint analysis. The constraints obtained from the GRB + BAO + $H(z)$ data are given in Tables 1 and 2. One-dimensional likelihood distributions and two-dimensional constraint contours are shown in blue in Figs.\ 5---10.

Amati relation parameters and intrinsic dispersion of the Amati relation determined using GRB + BAO + $H(z)$ data are model-independent and just a little different from the GRB-only values. These are listed in Table 2.

The values of the cosmological parameters determined from the GRB + BAO + $H(z)$ data do not differ significantly from those determined from the BAO + $H(z)$ data. In what follows we mention some interesting results from the joint analyses.

While the non-flat $\Lambda$CDM and XCDM cases results in $\Omega_{k0}$ values consistent with flat spatial hypersurfaces, the non-flat $\phi$CDM model favors closed geometry at 0.8$\sigma$. The Hubble constant values we measure are 2.3$\sigma$ to 3.1$\sigma$ lower than what is measured from the local expansion rate \citep{Riess2016}.

The flat and non-flat XCDM parametrizations favor dynamical dark energy density at 1.0$\sigma$ and 1.2$\sigma$ significance, respectively. The flat and non-flat $\phi$CDM model favor dynamical dark energy density at 1.1$\sigma$ and 1.5$\sigma$ significance, respectively.

From the $AIC$ values listed in Table 1, the most-favored model for the GRB + BAO + $H(z)$ data is the flat XCDM parametrization, and the least favored is the non-flat $\phi$CDM model, while the $BIC$ values most and least favor the flat $\Lambda$CDM model and the non-flat $\phi$CDM model.

\begin{figure*}
\begin{multicols}{2}
    \includegraphics[width=\linewidth]{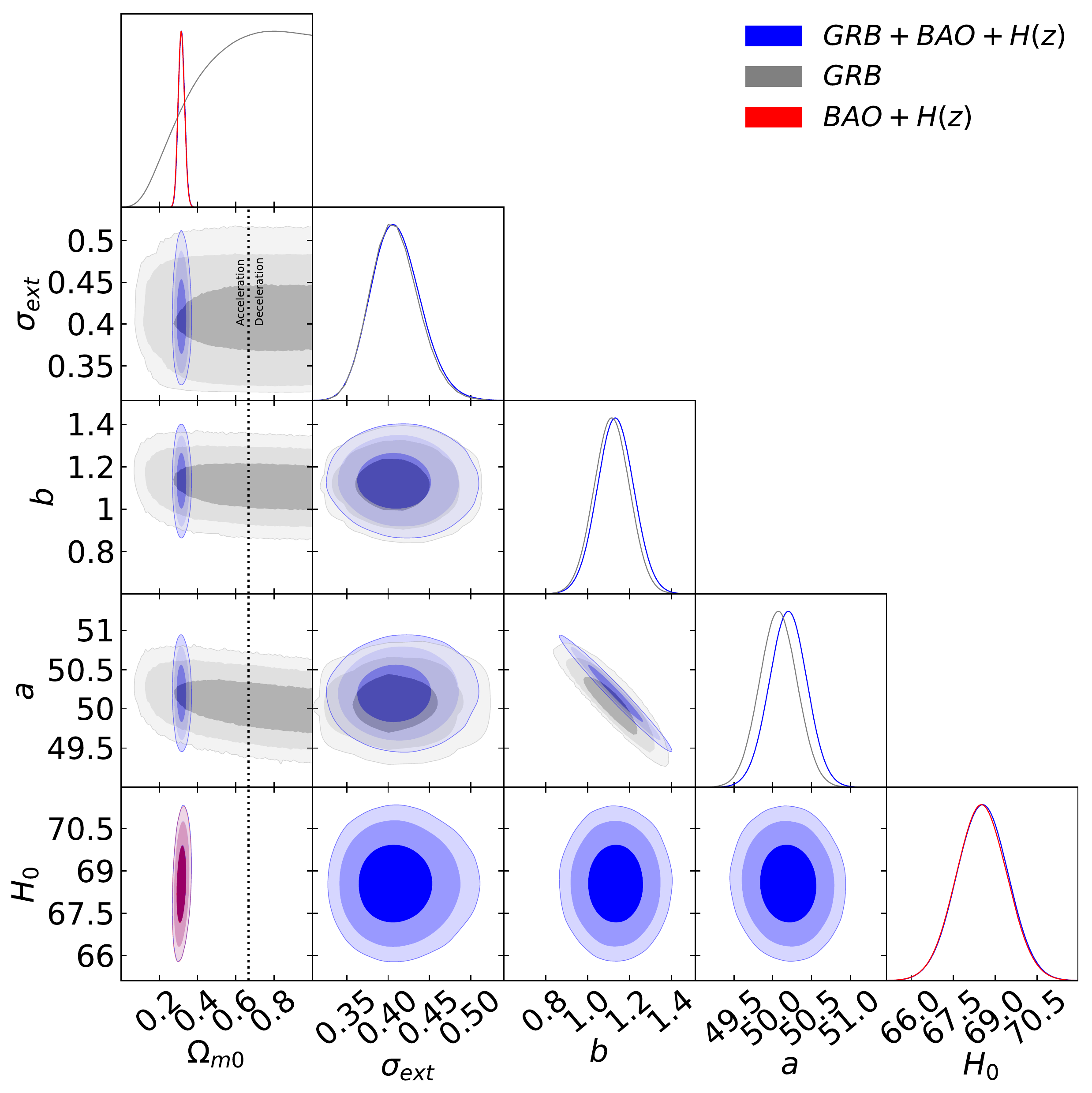}\par
    \includegraphics[width=\linewidth]{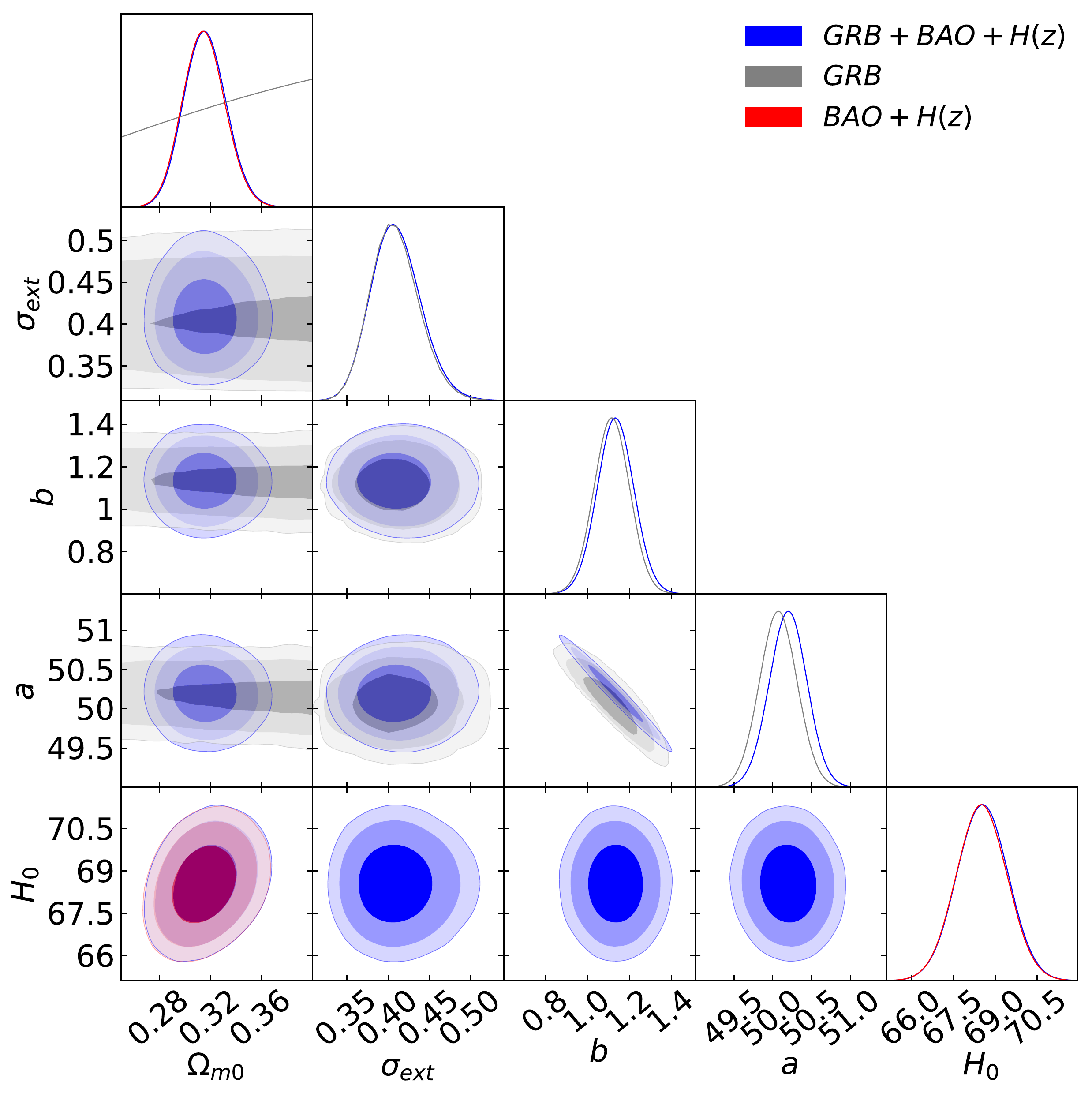}\par
\end{multicols}
\caption{Flat $\Lambda$CDM model one-dimensional likelihood distributions and two-dimensional contours at 1$\sigma$, 2$\sigma$, and 3$\sigma$ confidence levels using GRB (grey), BAO + $H(z)$ (red), and GRB + BAO + $H(z)$ (blue) data for all free parameters. The right panel shows the zoomed-in version of the left panel. Black dotted lines in the left panel are the zero acceleration line with currently accelerated cosmological expansion occurring to the left of the line.}
\label{fig:flat LCDM68 model with BAO, H(z) and QSO data}
\end{figure*}
%\begin{figure*}
%   \includegraphics[width=\linewidth]{FLCDMgrbbaohc2.pdf}\par
%\caption{FLCDM-GRB,BAO+H,GRB+BAO+H (zoom in version)}
%\label{fig:flat LCDM68 model with BAO, H(z) and QSO data}
%\end{figure*}

\begin{figure*}
\begin{multicols}{2}    
    \includegraphics[width=\linewidth]{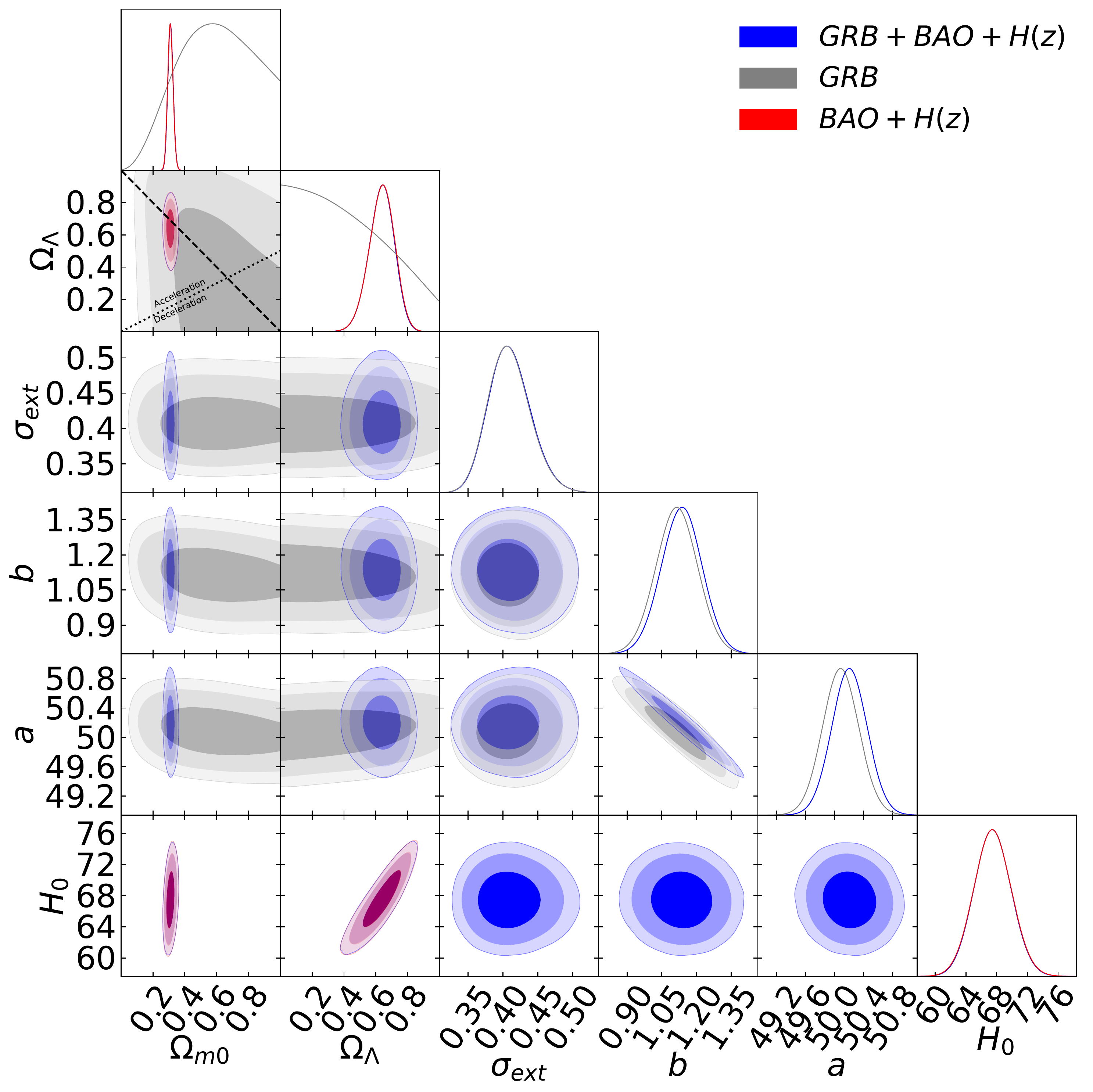}\par
    \includegraphics[width=\linewidth]{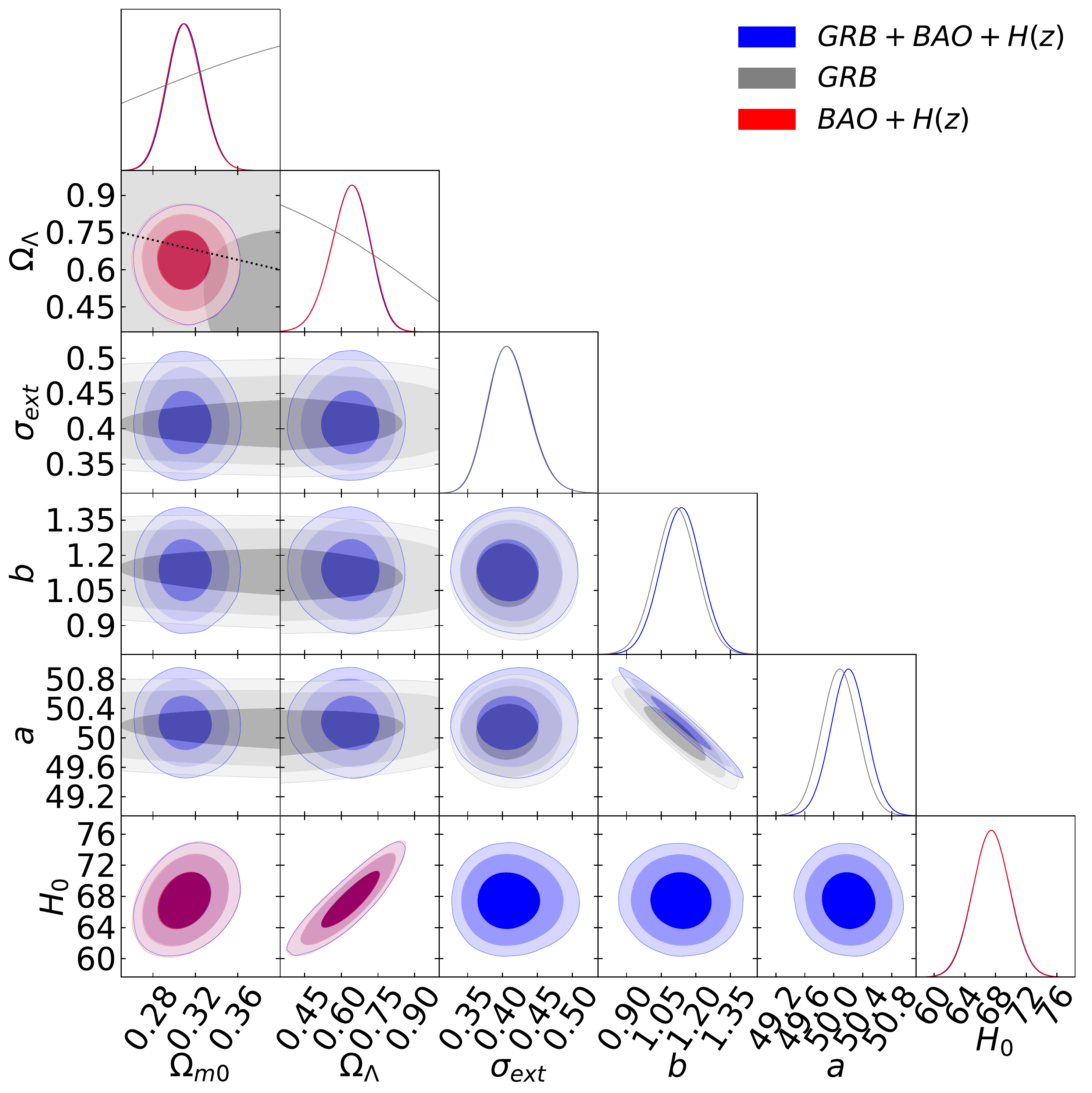}\par
\end{multicols}
\caption{Non-flat $\Lambda$CDM model one-dimensional likelihood distributions and two-dimensional contours at 1$\sigma$, 2$\sigma$, and 3$\sigma$ confidence levels using GRB (grey), BAO + $H(z)$ (red), and GRB + BAO + $H(z)$ (blue) data for all free parameters. The right panel shows the zoomed-in version of the left panel. The black dotted line in the $\Omega_{\Lambda}-\Omega_{m0}$ panel is the zero acceleration line with currently accelerated cosmological expansion occurring to the upper left of the line. The black dashed line in the $\Omega_{\Lambda}-\Omega_{m0}$ panel corresponds to the flat $\Lambda$CDM model, with closed hypersurface being to the upper right.}
\label{fig:flat LCDM68 model with BAO, H(z) and QSO data}
\end{figure*}
%\begin{figure*}
%    \includegraphics[width=\linewidth]{NLCDMgrbbaohc2.pdf}\par
%\caption{NLCDM-GRB,BAO+H,GRB+BAO+H (zoom in version)}
%\label{fig:flat LCDM68 model with BAO, H(z) and QSO data}
%\end{figure*}

\begin{figure*}
\begin{multicols}{2}
    \includegraphics[width=\linewidth]{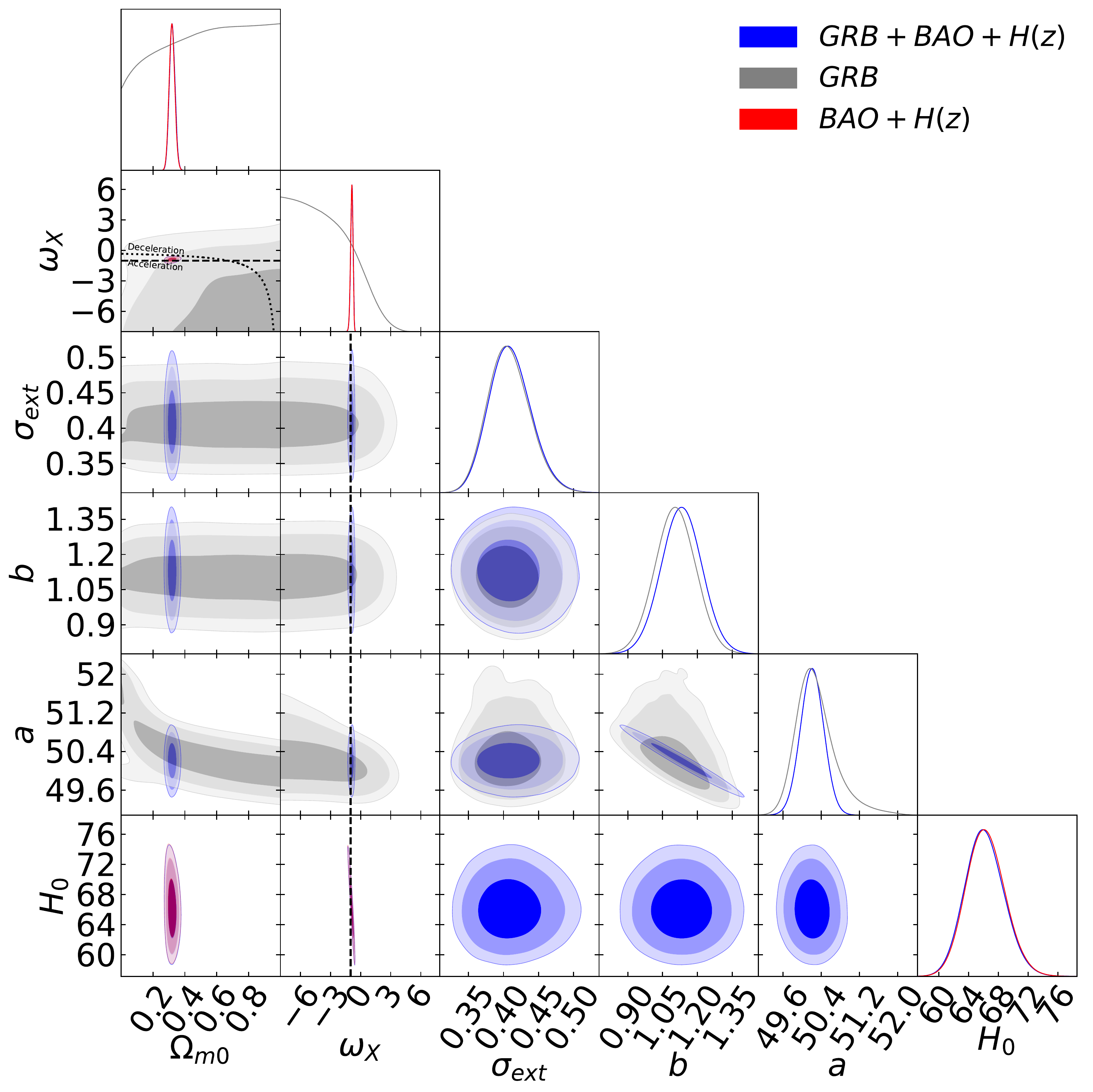}\par
    \includegraphics[width=\linewidth]{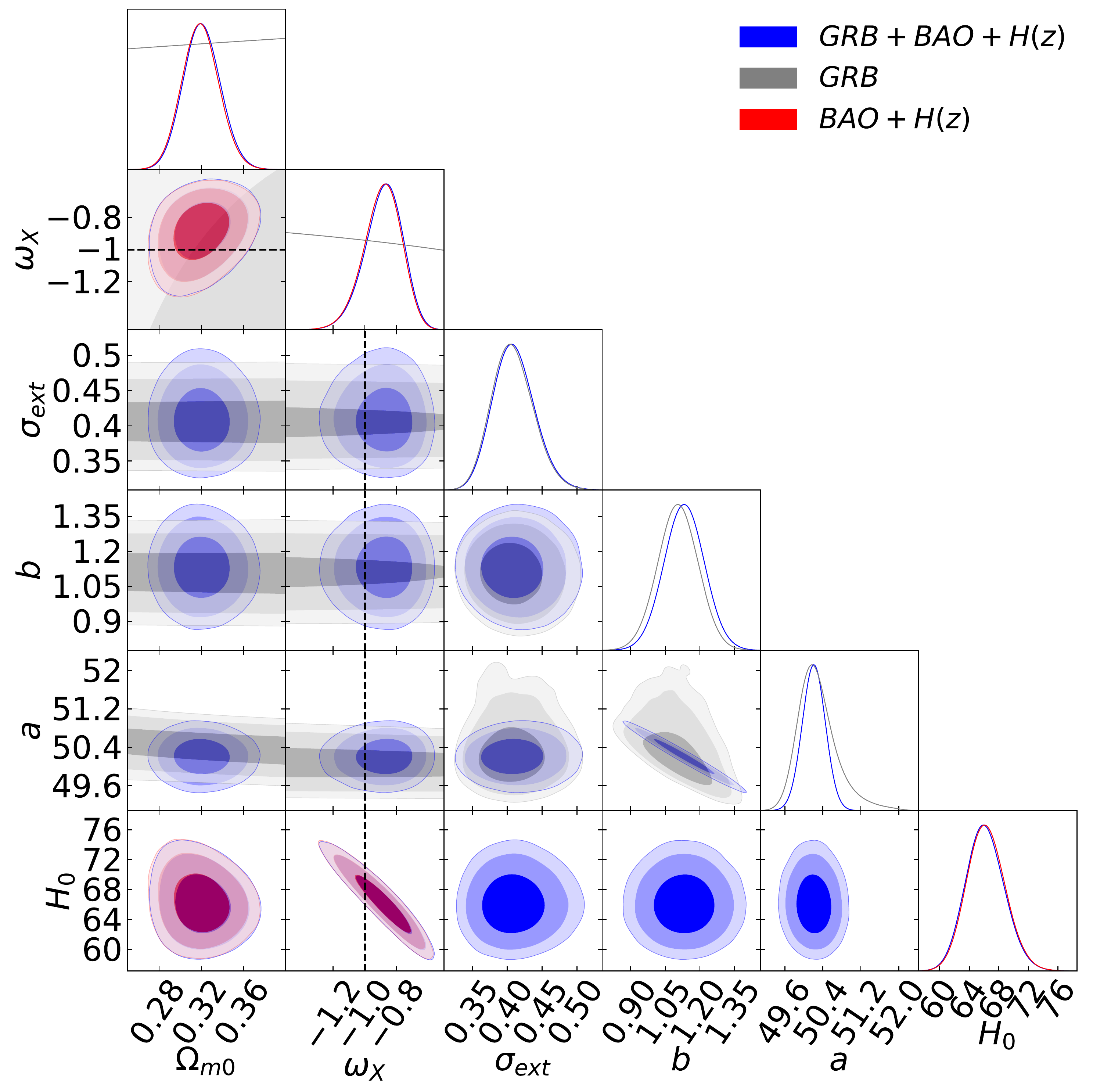}\par
\end{multicols}
\caption{Flat XCDM parametrization one-dimensional likelihood distributions and two-dimensional contours at 1$\sigma$, 2$\sigma$, and 3$\sigma$ confidence levels using GRB (grey), BAO + $H(z)$ (red), and GRB + BAO + $H(z)$ (blue) data for all free parameters. The right panel shows the zoomed-in version of the left panel. The black dotted line in the $\omega_X-\Omega_{m0}$ sub-panel of the left panel is the zero acceleration line  with currently accelerated cosmological expansion occurring below the line. The black dashed lines correspond to the $\omega_X = -1$ $\Lambda$CDM model.}
\label{fig:flat LCDM68 model with BAO, H(z) and QSO data}
\end{figure*}
%\begin{figure*}
%    \includegraphics[width=\linewidth]{FXCDMgrbbaohc2.pdf}\par
%\caption{FXCDM-GRB,BAO+H,GRB+BAO+H (zoom in version)}
%\label{fig:flat LCDM68 model with BAO, H(z) and QSO data}
%\end{figure*}

\begin{figure*}
\begin{multicols}{2}
    \includegraphics[width=\linewidth]{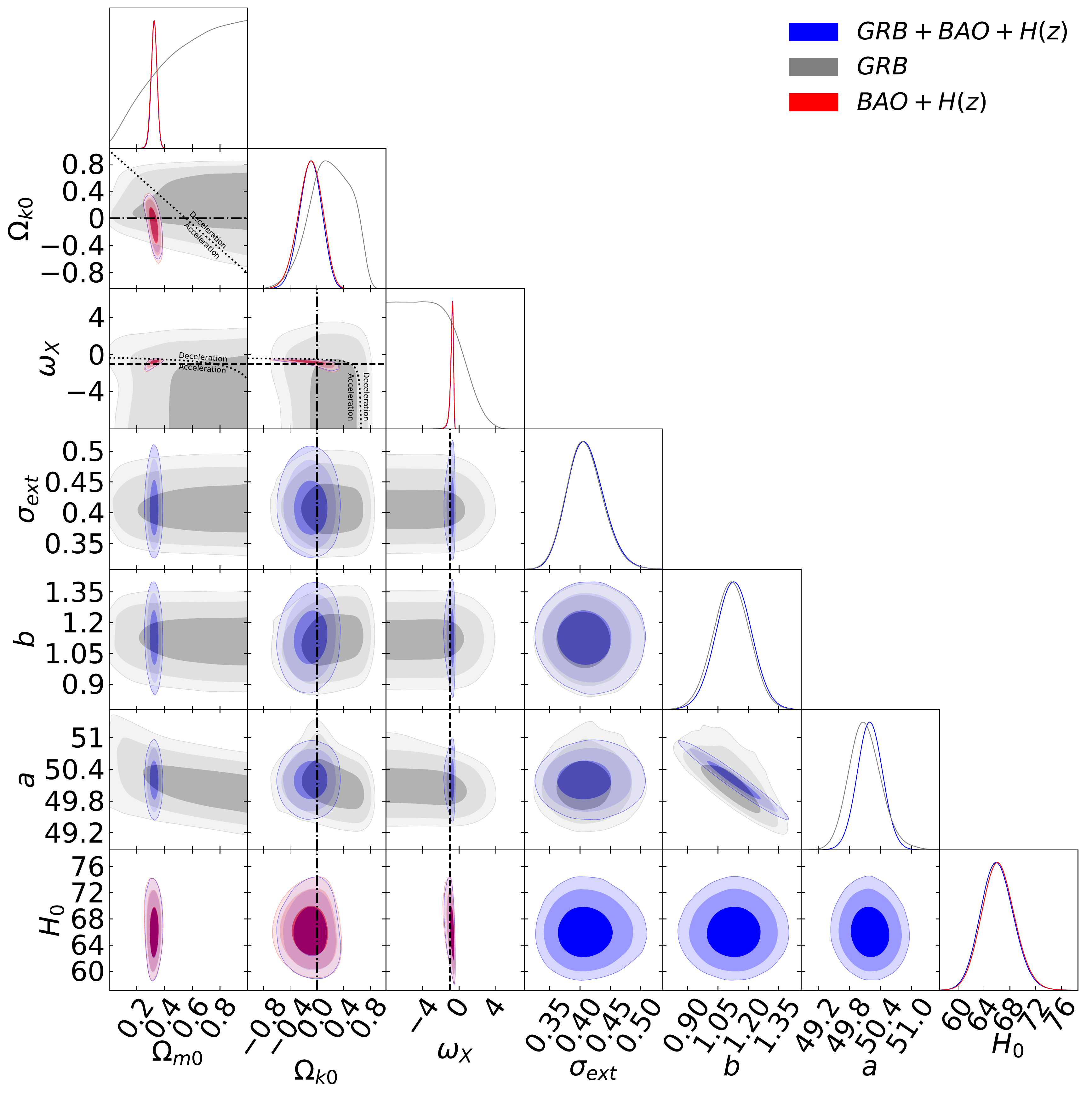}\par
    \includegraphics[width=\linewidth]{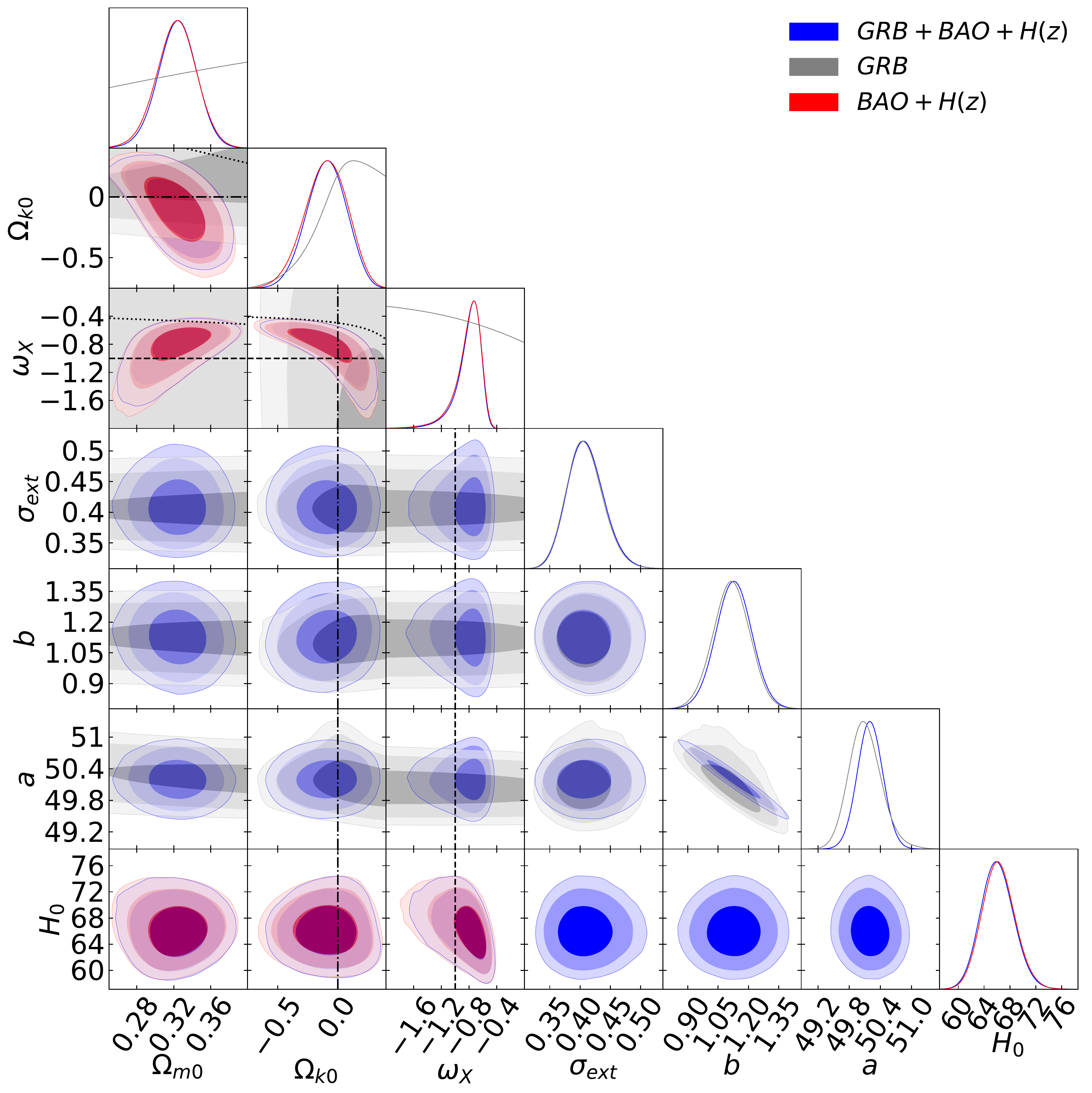}\par
\end{multicols}
\caption{Non-flat XCDM parametrization one-dimensional likelihood distributions and two-dimensional contours at 1$\sigma$, 2$\sigma$, and 3$\sigma$ confidence levels using GRB (grey), BAO + $H(z)$ (red), and GRB + BAO + $H(z)$ (blue) data for all free parameters. The right panel shows the zoomed-in version of the left panel. The black dotted lines in the $\Omega_{k0}-\Omega_{m0}$, $\omega_X-\Omega_{m0}$, and $\omega_X-\Omega_{k0}$ sub-panels of the left panel are the zero acceleration line with currently accelerated cosmological expansion occurring below the lines. Each of the three lines is computed with the third parameter set to the GRB + BAO + $H(z)$ data best-fit value of Table 1. The black dashed lines correspond to the $\omega_x = -1$ $\Lambda$CDM model. The black dot-dashed lines correspond to $\Omega_{k0} = 0$.}
\label{fig:flat LCDM68 model with BAO, H(z) and QSO data}
\end{figure*}
%\begin{figure*}
%    \includegraphics[width=\linewidth]{NXCDMgrbbaohc2.pdf}\par
%\caption{NXCDM-GRB,BAO+H,GRB+BAO+H (zoom in version)}
%\label{fig:flat LCDM68 model with BAO, H(z) and QSO data}
%\end{figure*}

\begin{figure*}
\begin{multicols}{2}
    \includegraphics[width=\linewidth]{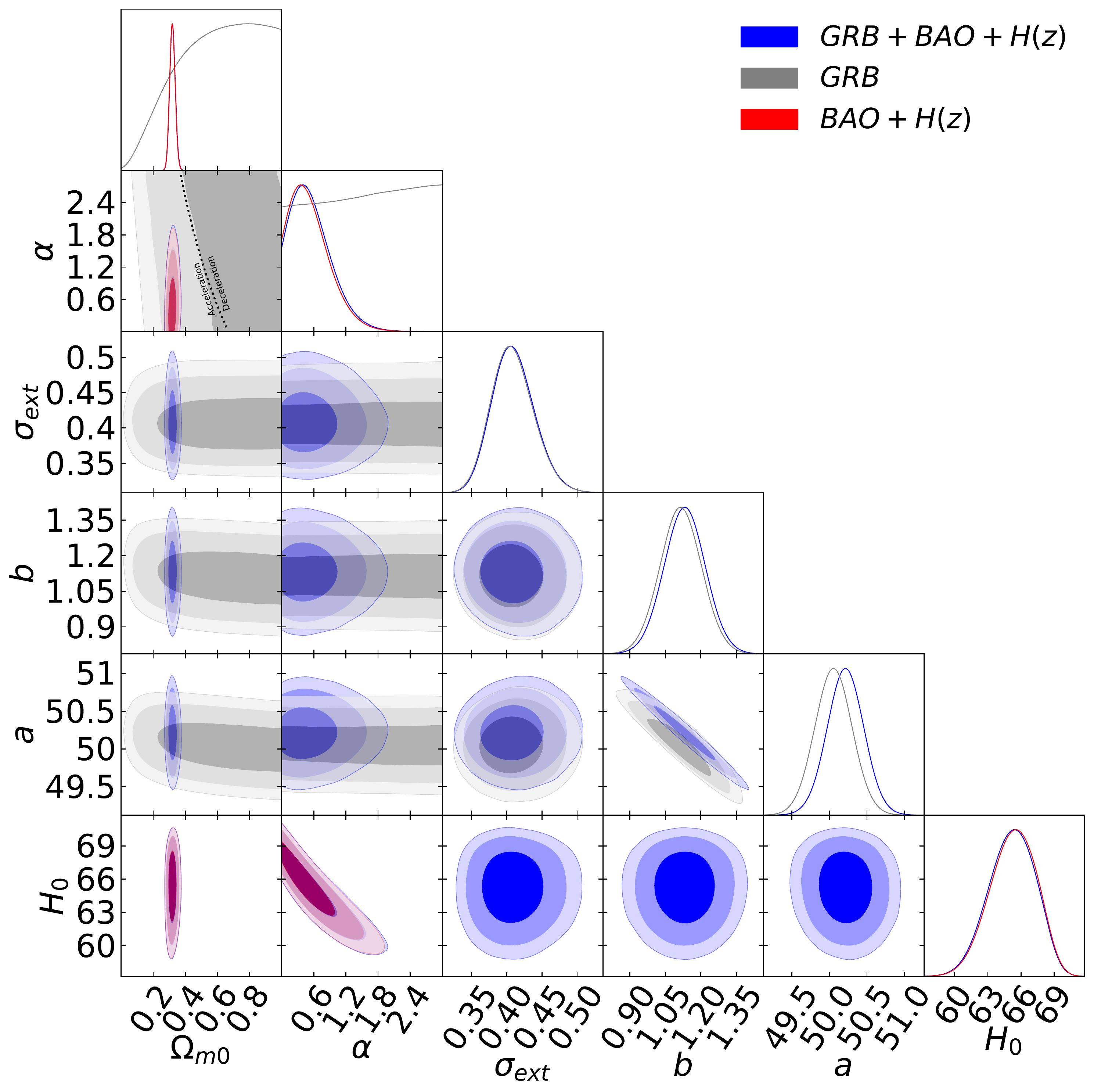}\par
    \includegraphics[width=\linewidth]{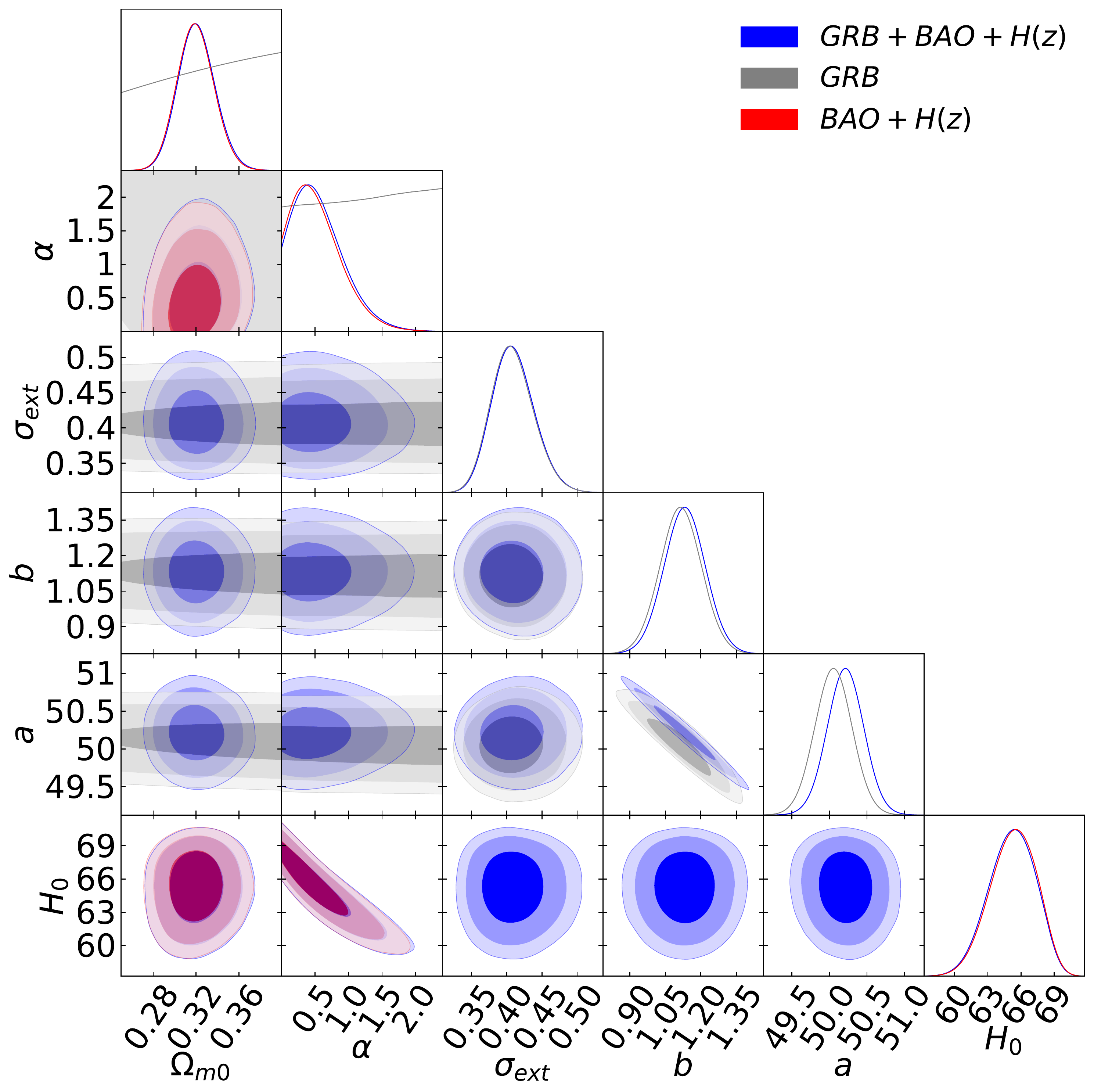}\par
\end{multicols}
\caption{Flat $\phi$CDM model one-dimensional likelihood distributions and two-dimensional contours at 1$\sigma$, 2$\sigma$, and 3$\sigma$ confidence levels using GRB (grey), BAO + $H(z)$ (red), and GRB + BAO + $H(z)$ (blue) data for all free parameters. The right panel shows the zoomed-in version of the left panel. The black dotted curved line in the $\alpha - \Omega_{m0}$ sub-panel of the left panel is the zero acceleration line with currently accelerated cosmological expansion occurring to the left of the line. The $\alpha = 0$ axis corresponds to the $\Lambda$CDM model.}
\label{fig:flat LCDM68 model with BAO, H(z) and QSO data}
\end{figure*}

%\begin{figure*}
%    \includegraphics[width=\linewidth]{fphiCDMgrbbaohc2.pdf}\par
%\caption{fphiCDM-GRB,BAO+H,GRB+BAO+H (zoom in version)}
%\label{fig:flat LCDM68 model with BAO, H(z) and QSO data}
%\end{figure*}

\begin{figure*}
\begin{multicols}{2}
    \includegraphics[width=\linewidth]{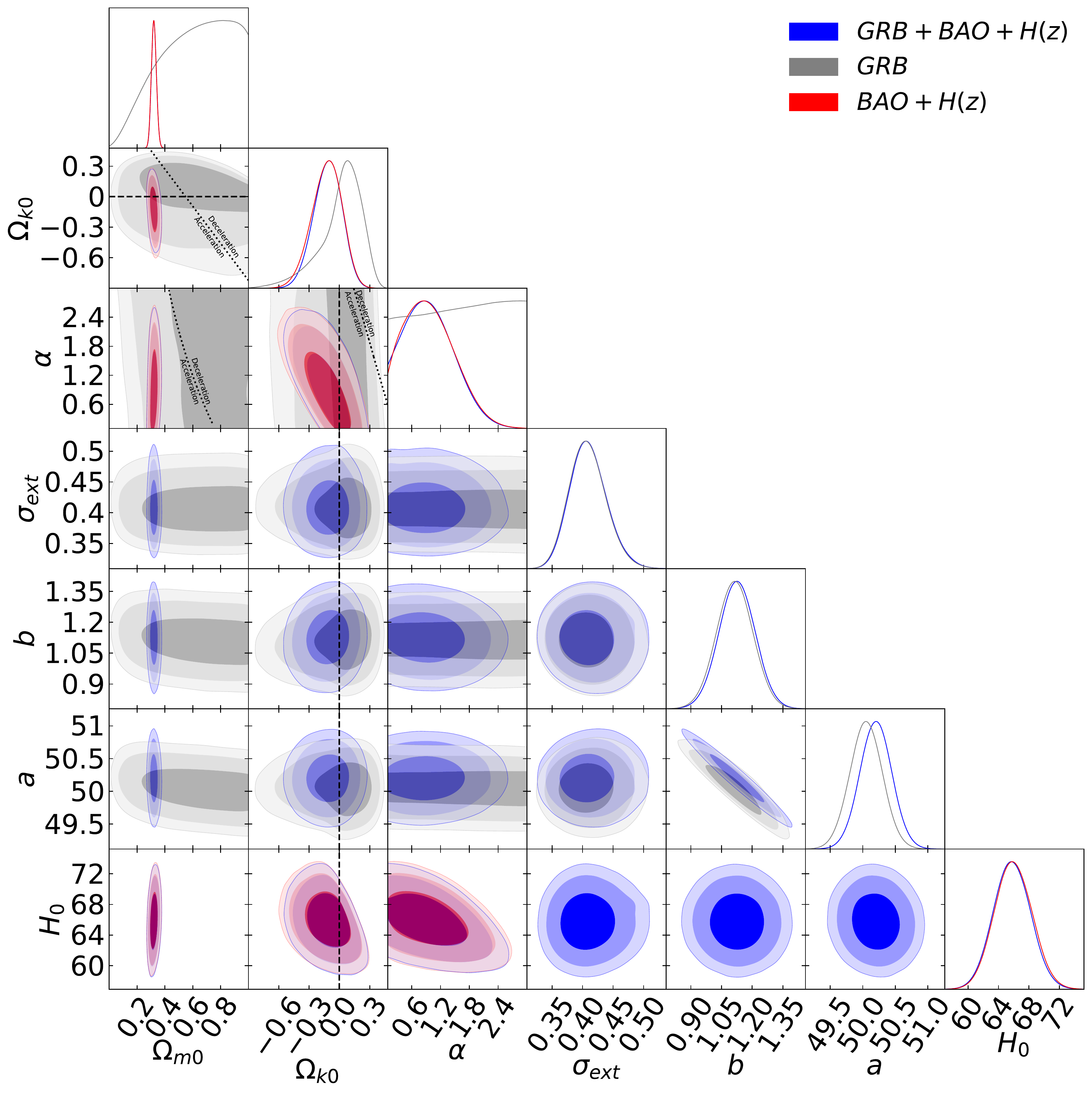}\par
    \includegraphics[width=\linewidth]{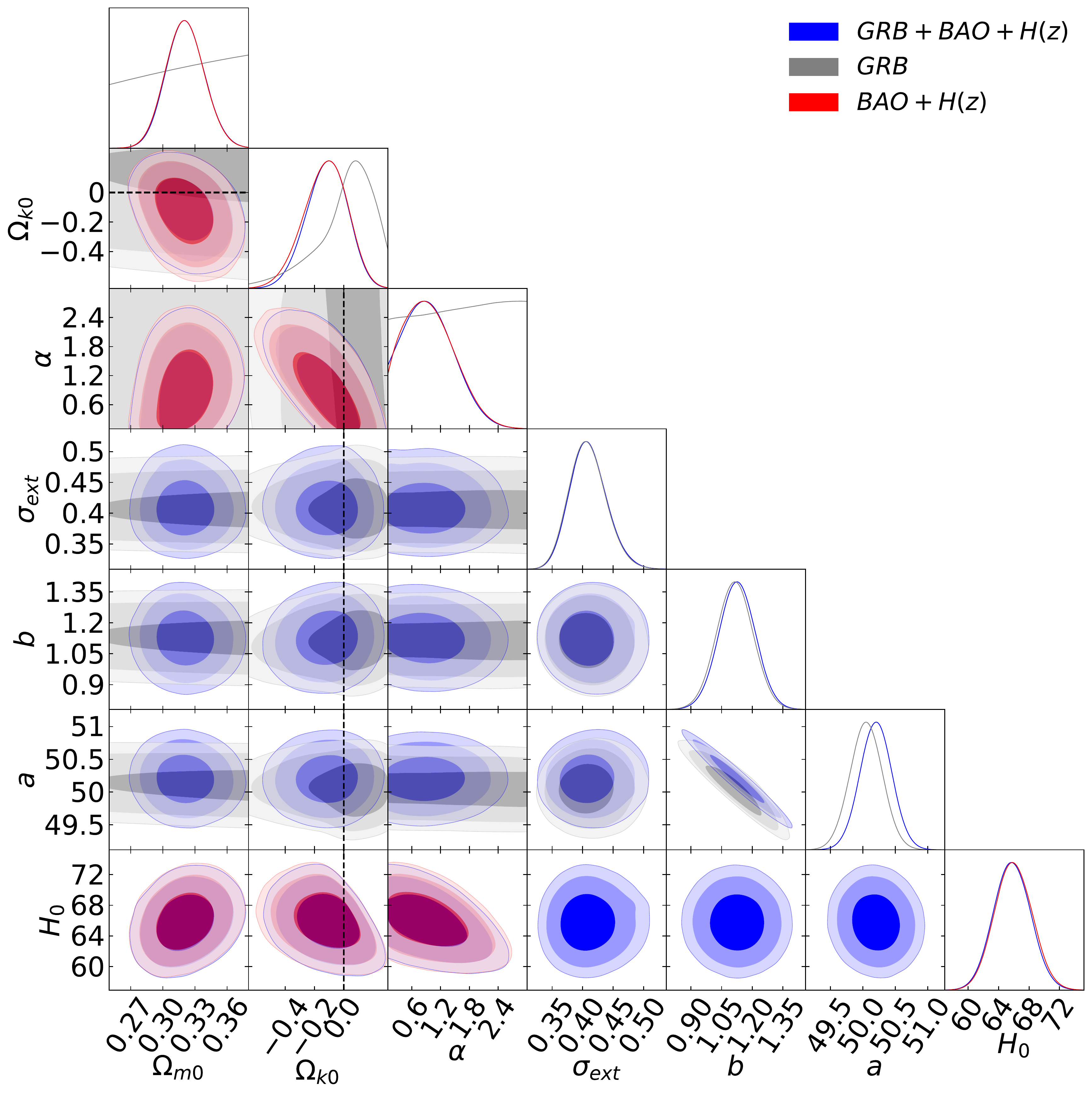}\par
\end{multicols}
\caption{Non-flat $\phi$CDM model one-dimensional likelihood distributions and two-dimensional contours at 1$\sigma$, 2$\sigma$, and 3$\sigma$ confidence levels using GRB (grey), BAO + $H(z)$ (red), and GRB + BAO + $H(z)$ (blue) data for all free parameters. The right panel shows the zoomed-in version of the left panel. The black dotted lines in the $\Omega_{k0}-\Omega_{m0}$, $\alpha-\Omega_{m0}$, and $\alpha-\Omega_{k0}$ sub-panels of the left panel are the zero acceleration lines with currently accelerated cosmological expansion occurring below the lines. Each of the three lines is computed with the third parameter set to the GRB + BAO + $H(z)$ data best-fit value of Table 1. The $\alpha = 0$ axis corresponds to the $\Lambda$CDM model. The black dashed straight lines correspond to $\Omega_{k0} = 0$.}
\label{fig:flat LCDM68 model with BAO, H(z) and QSO data}
\end{figure*}

\section{CONCLUSION}
From the analysis of the combined GRB data in six different cosmological models, we find that the Amati relation is independent of the cosmological model. This is the most comprehensive demonstration to date of this model independence, and shows that these GRBs can be standardized and used to derive cosmological constraints.

However, even the joint GRB measurements have large uncertainty and so cosmological constraints obtained from them are not so restrictive. They are mostly only able to set a lower limit on the non-relativistic matter density parameter $(\Omega_{m0})$ but are a little more successful at setting (weak) limits on the spatial curvature density parameter $(\Omega_{k0})$. They can only set an upper limit on the cosmological constant energy density parameter $(\Omega_{\Lambda})$ and on $\omega_X$ in the XCDM parametrization, but are unable to constrain $\alpha$ in the $\phi$CDM model.

We note that in many previous analyses cosmological constraints have been obtained using GRB data with fixed Amati relation parameters (fixed using additional external  information), or with a fixed value of $\Omega_{m0}$, or calibrated using external calibrator (such as Type Ia supernovae). Such constraints are tighter than what we have determined here, but are not purely GRB constraints. As this is still a developing area of research, we believe it is important to also examine GRB-only constraints, as we have done here. In addition, since we simultaneously fit all the cosmological parameters and the Amati relation parameters our results are free of the circularity problem but are less constraining.

Current GRB data are not able to constrain cosmological parameters very restrictively but future improved GRB data should provide more restrictive constraints and help study the largely unexplored $z \sim 2-\textbf{10}$ part of the universe.

\section{ACKNOWLEDGEMENTS}
We thank Fayin Wang, Lado Samushia, Javier De Cruz, Joe Ryan, and Shulei Cao for useful discussions. We are grateful to the Beocat Research Cluster at Kansas State University team. This research was supported in part by DOE grants DE-SC0019038 and DE-SC0011840.

\section*{Data availability}
The data underlying this article are publicly available in
\cite{Dirisa2019}.



%\begin{figure*}
%    \includegraphics[width=\linewidth]{nphiCDMgrbC.pdf}\par
%\caption{nphiCDM-GRBS}
%\label{fig:flat LCDM68 model with BAO, H(z) and QSO data}
%\end{figure*}

%\begin{figure*}
%    \includegraphics[width=\linewidth]{nphiCDMgrbbaohc2.pdf}\par
%\caption{nphiCDM-GRB,BAO+H,GRB+BAO+H (zoom in version)}
%\label{fig:flat LCDM68 model with BAO, H(z) and QSO data}
%\end{figure*}

%%%%%%%%%%%%%%%%%%%%%%%%%%%%%%%%%%%%%%%%%%%%%%%%%%

%%%%%%%%%%%%%%%%% APPENDICES %%%%%%%%%%%%%%%%%%%%%
\begin{comment}
\appendix

\newpage~

\newpage~

\newpage~

\newpage~

\newpage~

\newpage~

\newpage~

\newpage~

\newpage~

\newpage~

\end{comment}

%%%%%%%%%%%

% Don't change these lines
\bsp	% typesetting comment
\label{lastpage}
\end{document}